\pdfoutput=1
\documentclass[ijds,sglanonrev]{informs4}

\usepackage{eqndefns-left}
\RequirePackage{tgtermes}
\RequirePackage{newtxtext}
\RequirePackage{newtxmath}
\RequirePackage{bm}
\RequirePackage{endnotes}

\OneAndAHalfSpacedXII

\usepackage{amsmath}
\usepackage{graphicx}
\usepackage{algorithm}
\usepackage{algorithmic}
\usepackage{subfig}
\usepackage{multirow}
\usepackage{booktabs}
\usepackage{longtable}
\usepackage{comment}
\usepackage{enumitem}

\usepackage{natbib}
\bibpunct[, ]{(}{)}{,}{a}{}{,}%
\def\bibfont{\small}%
\EquationsNumberedThrough
\TheoremsNumberedThrough
\ECRepeatTheorems

\MANUSCRIPTNO{}

\usepackage[
    colorlinks=true,
    linkcolor=blue,
    citecolor=blue,
    urlcolor=blue
]{hyperref}

\newcommand{\FANCI}{\textsc{FANCI}}

\newcommand{\FD}{\operatorname{FD}}

\newcommand{\FANCIFull}{\textsc{FANCI-Full}}
\newcommand{\FANCIGeo}{\textsc{FANCI-Geo}}
\newcommand{\FANCIUnc}{\textsc{FANCI-Unc}}
\newcommand{\MMC}{\textsc{MMC}}
\newcommand{\RAND}{\textsc{RAND}}

\begin{document}

\RUNAUTHOR{Liu et al.}
\RUNTITLE{FANCI under Network Interference}

\TITLE{Fill-Distance-Guided Active Learning for Causal Function Recovery under Network Interference}

\ARTICLEAUTHORS{%
\AUTHOR{Peiyao Liu\textsuperscript{1}, Hongtao Zhu\textsuperscript{1}, Sizhe Zhang\textsuperscript{2}, Yang Su\textsuperscript{3}, Wesley Zhao\textsuperscript{4}, Nan Chen\textsuperscript{1}}
\AFF{\textsuperscript{1}Department of Industrial Systems Engineering and Management,
National University of Singapore, Singapore 117576\\
\textsuperscript{2}Duotech, Singapore\quad
\textsuperscript{3}Apple, United States\quad
\textsuperscript{4}Tencent, Palo Alto, CA, United States\\
\EMAIL{liupy25@nus.edu.sg}, \EMAIL{hongtao.zhu@u.nus.edu}, \EMAIL{zhangsizhe1987@gmail.com},\\
\EMAIL{suyang1025@gmail.com}, \EMAIL{zzy287@gmail.com}, \EMAIL{isecn@nus.edu.sg}}
}

\ABSTRACT{%
Causal inference seeks to learn how outcomes change under alternative interventions, but network interactions complicate this task. Under network interference, outcomes depend on both own treatment and neighbors' treatments, making causal effects functions of neighborhood exposure rather than single treatment contrasts. Existing methods primarily target prespecified causal contrasts or their estimation precision, leaving global response-function recovery largely unexplored.
We study sequential experimentation under a limited experimental budget, with the goal of recovering the full exposure-dependent population-average response function. To this end, we propose FANCI (\underline{F}ill-Distance-Guided \underline{A}ctive Learning for \underline{N}etwork-Interference \underline{C}ausal \underline{I}nference). For function learning, treatment-specific Gaussian processes flexibly model nonlinear conditional response surfaces. For experimental design, we formulate global response-function recovery as a coverage problem and propose a fill-distance criterion that guides new experiments toward poorly covered regions. To account for network constraints on attainable exposure patterns, we develop full-network and target-screened strategies for selecting feasible network--assignment pairs.
Theoretically, we derive uniform error bounds linking fill distance to response-function estimation and characterize the approximation loss from target screening. Simulations and an online-game experiment show that FANCI improves recovery of nonlinear exposure-dependent effects and reveals treatment patterns obscured when network interference is ignored.

}


\maketitle

%%%%%%%%%%%%%%%%%%%%%%%%%%%%%%%%%%%%%%%%%%%%%%%%%%%%%%%%%%%%%%%%%%%%%%

\section{Introduction}
\label{sec:introduction}

Causal inference asks how an outcome would change under alternative
interventions. For individual $i$, let $A_i\in\{0,1\}$ denote a binary
treatment and $Y_i(a)$ the potential outcome under treatment level $a$.
The basic causal question is to compare $Y_i(1)$ and $Y_i(0)$, although
only the outcome under the treatment actually received can be observed
\citep{fernandez2022causal}. Such questions arise in many applications.
In public health, causal methods are used to evaluate medical and policy
interventions \citep{halloran1995causal}. In education, they assess the
effects of programs and policies \citep{hong2006evaluating}. On online
platforms, randomized experiments are widely used to evaluate product
and incentive interventions \citep{eckles2016estimating}.

\begin{figure}[htbp]
    \centering
    \includegraphics[width=0.8\linewidth]{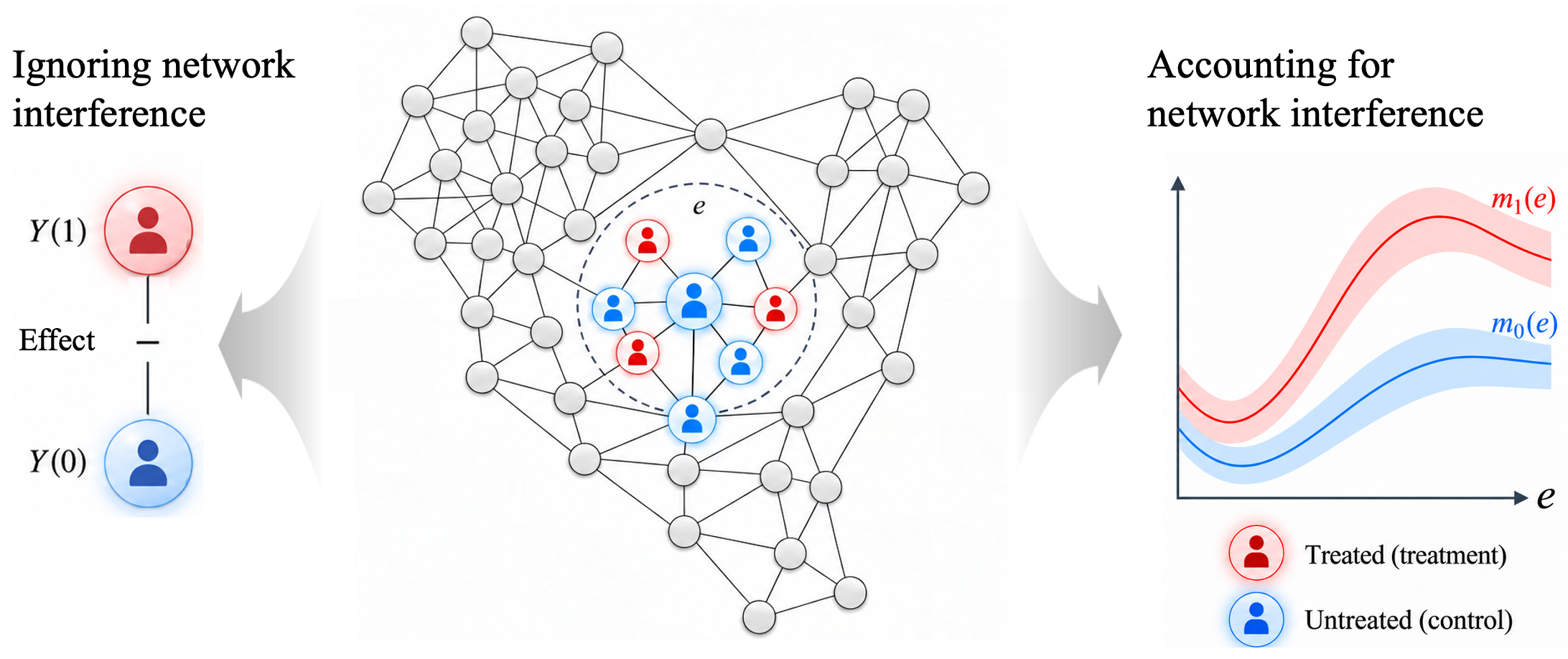}
    \caption{From Scalar Causal Effects to Exposure-Dependent Response Functions under Network Interference}
    \label{fig:network-interference}
\end{figure}

Standard causal analyses often assume that an individual's outcome is
affected only by the treatment that individual receives
\citep{rubin1980randomization}. However, this assumption may fail when
individuals are connected through social, communication, or operational
networks. For example, vaccinating one person may change the infection
risk of untreated contacts \citep{hudgens2008toward}, while an
intervention delivered to one online user may affect the behavior of
connected peers \citep{aronow2017estimating}. Figure~\ref{fig:network-interference}
illustrates this issue in an online-game setting, where a player's
engagement may depend on both the player's own treatment and treatments
received by in-game friends. This dependence is known as
\emph{network interference}. Let $e$ denote neighborhood exposure.
The relevant potential outcome can then be written as $Y_i(a,e)$
\citep{forastiere2021identification}, and the corresponding
population-average response function is $m_a(e)=\mathbb{E}[Y_i(a,e)]$,
averaging over pretreatment covariates. Under network interference,
these response functions may vary with $e$, and treatment-effect
contrasts may therefore also depend on neighborhood exposure.

A growing literature has developed methods for identifying and
estimating causal effects under network interference
\citep{aronow2017estimating,liu2016inverse,park2022efficient}.
However, much of this literature focuses on causal contrasts at
prespecified treatment and exposure levels rather than recovering
the full response functions. In particular, $m_0(e)$ and $m_1(e)$
characterize population-average outcomes across neighborhood
exposure levels under control and treatment, respectively.
These functions may exhibit nonlinear or nonmonotone patterns,
with important features such as thresholds, saturation, or
interior extrema that a few prespecified contrasts could miss.
Our first goal is therefore to recover both population-average
response functions over the exposure range of interest from
the data collected thus far.

Another challenge is deciding where to collect additional data to recover
these functions efficiently. This is particularly important when experiments
are costly \citep{huang2026active,kosolwattana2025online}. For example,
an online-game platform may contain hundreds of networks but be able to
experiment on only a small subset. While existing sequential designs
primarily target prespecified causal effects
\citep{viviano2026experimental,zhang2025online}, recovering full response
functions requires experiments that improve learning across the exposure
domain. This calls for a criterion that evaluates each candidate experiment's
contribution to global function recovery. Under network interference,
however, such design is challenging because neighborhood exposures cannot
be chosen independently: a single treatment assignment simultaneously
determines the exposures of connected individuals, and different networks
may generate different exposure patterns. Selecting the next experiment
therefore requires jointly choosing a feasible network and its treatment
assignment.

Motivated by these challenges, we develop FANCI
(\underline{F}ill-Distance-Guided \underline{A}ctive Learning for
\underline{N}etwork-Interference \underline{C}ausal
\underline{I}nference), a unified framework for learning the full
population-average response function through sequential network experiments.
For function estimation, we employ treatment-specific Gaussian processes (GP)
to flexibly learn nonlinear conditional response surfaces and then average these surfaces over the target covariate distribution to obtain the population-average response functions $m_a(e)$. For experimental design,
we view global response-function recovery as a coverage problem and propose a
fill-distance criterion to quantify how well the current data cover the
target domain. The next experiment is then selected to improve this
coverage through a feasible network--assignment pair. In this way,
function estimation and data collection are directly aligned with the
goal of recovering the full population-average response function.

The paper makes four contributions:
\begin{itemize}

\item We introduce the problem of sequential global response-function
recovery under network interference and, to the best of our knowledge,
provide its first rigorous formulation.

\item We propose a coverage-based acquisition criterion that translates
global response-function recovery into an experimental-design objective
over feasible network-induced observations.

\item We develop full-network and target-screened optimization strategies
for selecting network--assignment pairs, with the latter reducing
computational cost through candidate screening.

\item We establish uniform error bounds linking fill distance to
response-function estimation error and characterize the approximation
loss introduced by target screening.

\end{itemize}

The remainder of the paper is organized as follows.
Section~\ref{sec:related-literature} reviews the related literature.
Section~\ref{sec:causal-framework} presents the causal framework,
estimands, and sequential design problem.
Section~\ref{sec:methodology} develops the \FANCI\ methodology.
Section~\ref{sec:theory} provides the theoretical analysis.
Section~\ref{sec:numerical} reports the simulation and online-game
studies. Section~\ref{sec:conclusion} concludes the paper.
%%%%%%%%%%%%%%%%%%%%%%%%%%%%%%%%%%%%%%%%%%%%%%%%%%%%%%%%%%%%%%%%%%%%%%
\section{Related Work}
\label{sec:related-literature}

In this section, we review the literature on causal inference under interference and the
experimental designs developed to learn causal effects in such settings.

\subsection{Causal Inference under Interference}
Under the potential-outcomes framework,
a binary treatment is associated with two potential outcomes,
$Y_i(1)$ and $Y_i(0)$, and a population causal effect is typically
defined through contrasts such as
$\mathbb{E}[Y_i(1)-Y_i(0)]$
\citep{neyman1990application,rubin1974estimating}.
A large literature has developed methods for estimating such effects,
including regression adjustment and matching
\citep{rubin1979multivariate}, propensity-score methods
\citep{rosenbaum1983central}, and weighting and semiparametric
estimation \citep{hahn1998role,hirano2003efficient}.
These classical formulations assume that an individual's potential
outcome depends only on that individual's own treatment. However, this assumption
may fail when individuals are connected through social, communication,
or other interaction networks, because treatments assigned to connected
individuals may also affect the outcome of interest.

Under such network interference, defining a causal effect becomes more
challenging because both an individual's own treatment and the treatment
environment created by other individuals must be specified.
To address this issue, \citet{sobel2006randomized} studied treatment effects in the presence of
spillovers, while \citet{hudgens2008toward} formalized direct, indirect,
total, and overall effects under partial interference. \citet{hu2022average} studied average direct and indirect
effects under general interference. 
For more general network settings, \citet{aronow2017estimating}
introduced exposure mappings that summarize treatments received by
other units through an individual-specific exposure $e$.
Writing $m_a(e)=\mathbb{E}[Y_i(a,e)],$
 the direct effect at exposure $e$ compares $m_1(e)$ and $m_0(e)$,
whereas a spillover effect compares $m_a(e)$ and $m_a(e')$ across
different exposure levels. 

A substantial literature has developed methods for estimating causal effects under network interference. For instance, 
\citet{aronow2017estimating} developed randomization-based estimators
for average potential outcomes under specified exposure conditions and
their contrasts. \citet{liu2016inverse} proposed
inverse-probability-weighted and stabilized estimators for direct,
indirect, total, and overall effects under specified treatment-allocation
strategies. \citet{park2022efficient} derived semiparametric efficiency
bounds and efficient estimators for causal effects under partial
interference, while \citet{li2022random} established asymptotic
properties for average direct and indirect effect estimators as the
network grows. When exposure varies continuously, $m_a(e)$ and the corresponding causal effects become functions of $e$. However, these studies do not primarily aim to recover the entire response functions over the exposure domain.

Moving beyond estimation at prespecified exposure levels, \citet{forastiere2021identification} used generalized propensity-score adjustment to estimate average responses across neighborhood exposure levels under network interference. \citet{forastiere2022estimating} further developed a Bayesian approach combining propensity-score adjustment with spline-based outcome models to capture nonlinear variation in these responses.  Inspired by these studies, we aim to recover population-average response functions over the entire exposure domain. Our approach differs from these methods by employing treatment-specific GPs to flexibly model conditional response surfaces in a nonparametric manner, while further integrating function estimation with sequential experimental design to improve global recovery of population-average response functions.

\subsection{Experimental Design under Interference}

Experimental design chooses treatment assignments to obtain informative
observations. Under network interference, an individual's treatment
assignment may also affect the neighborhood exposures of others, so
informative designs must account for both. \citet{baird2018optimal}
studied randomized saturation designs that vary treatment proportions
across groups to improve the power for detecting treatment and
spillover effects. When the behavior generating spillovers cannot be
directly assigned, \citet{eckles2016estimating} proposed
peer-encouragement designs that use randomized encouragement to
create exogenous variation in peer behavior. These designs account
for interference when constructing informative experiments, but do
not primarily use detailed network topology to determine treatment
assignments.

When the interaction network is observed, its topology can inform
treatment allocation because assignments jointly determine the
exposures of connected individuals. \citet{ugander2013graph} proposed
graph-cluster randomization, while \citet{jagadeesan2020designs}
used graph quasi-coloring to construct informative treatment and
exposure configurations. For network A/B testing,
\citet{zhang2022locally} developed locally optimal designs under
network-correlated outcomes, \citet{zhang2025rerandomization}
proposed rerandomization designs under network dependence or
interference, and \citet{bui2026genetic} optimized a Bayesian design
criterion using genetic algorithms. These methods exploit network
structure to improve the estimation of prespecified treatment effects,
but do not target recovery of entire exposure-dependent response
functions. Moreover, the designs are generally specified before
experimental outcomes are observed.

More recent work allows experimental designs to adapt as data are
collected. \citet{viviano2026experimental} developed a two-wave
network experiment in which pilot outcomes guide participant
selection and treatment assignment in the second wave.
\citet{zhang2025online} considered fully sequential design under
network interference, balancing treatment-effect estimation accuracy
and cumulative regret. Beyond causal inference, sequential designs
for expensive experiments also target global response-surface
learning. In particular, \citet{alaeddini2019sequential} proposed
a Laplacian-regularized V-optimal criterion, while GP-based active
learning uses criteria such as posterior uncertainty or model change
\citep{zhao2023batch}. Despite these advances, sequential network
designs mainly target prespecified causal effects, whereas general
response-surface designs allow more flexible input selection.
Sequential recovery of entire causal response functions under
network-constrained experimentation therefore remains largely
unexplored.

\section{Problem Formulation}
\label{sec:causal-framework}

This section establishes the causal framework, causal estimands, and
sequential experimental-design problem under network interference. Throughout the paper,
uppercase letters denote random quantities, whereas lowercase letters
denote their realizations. Bold symbols denote vectors or matrices.

\subsection{Causal Framework under Network Interference}
\label{sec:causal-framework-interference}

We consider a finite collection of candidate networks indexed by
\(\mathcal K=\{1,\ldots,K\}\). Network \(k\) contains \(n_k\)
individuals, with node set \(\mathcal V_k=\{1,\ldots,n_k\}\), and is
represented by a known weighted adjacency matrix
\(\boldsymbol W_k=[w_{kij}]_{n_k\times n_k}
\in\mathbb R_{+}^{n_k\times n_k}\). We set \(w_{kii}=0\) and interpret
\(w_{kij}>0\) as a relationship from individual \(j\) to individual
\(i\). The unweighted case corresponds to \(w_{kij}\in\{0,1\}\).

We assume each candidate network acts as an experimental cluster: individuals may affect others within the same network, but different candidate networks do not interfere with one another
\citep{hudgens2008toward}.
 For individual \(i\) in network
\(k\), let
\(\mathcal N_{ki}=\{j\in\mathcal V_k:w_{kij}>0\}\)
denote its neighborhood and let
\(d_{ki}=|\mathcal N_{ki}|\). The random assignment vector is
\(\boldsymbol A_k=(A_{k1},\ldots,A_{kn_k})^\top\), with realization
\(\boldsymbol a_k=(a_{k1},\ldots,a_{kn_k})^\top\). We write
\(\boldsymbol A_{k[\mathcal N_{ki}]}
=(A_{kj}:j\in\mathcal N_{ki})^\top\)
for the neighboring treatments and define
\(\boldsymbol a_{k[\mathcal N_{ki}]}\in \{0,1\}^{d_{ki}}\) analogously. Since this vector
may influence the outcome through network interference and its
dimension varies across individuals, we summarize it through a common
exposure mapping.

\begin{definition}[Neighborhood exposure mapping]
\label{def:exposure-mapping}

Let $\phi:
    \bigcup_{d\geq1}
    (
        \{0,1\}^{d}\times\mathbb R_{+}^{d}
    )
    \rightarrow
    \mathcal E
    \subseteq\mathbb R$
be a prespecified mapping that converts neighboring treatments and
their relationship weights into a scalar exposure. For individual
\(i\) in network \(k\), let
\(\boldsymbol w_{ki}
=(w_{kij}:j\in\mathcal N_{ki})^\top
\in\mathbb R_{+}^{d_{ki}}\)
collect the weights linking its neighbors to individual \(i\). The
random neighborhood exposure and its realization under
\(\boldsymbol a_k\) are
\[
    E_{ki}
    =
    \phi
    \left(
        \boldsymbol A_{k[\mathcal N_{ki}]};
        \boldsymbol w_{ki}
    \right),
    \qquad
    e_{ki}
    =
    \phi
    \left(
        \boldsymbol a_{k[\mathcal N_{ki}]};
        \boldsymbol w_{ki}
    \right).
\]
\end{definition}

The exposure mapping \(\phi\) may be defined using the number or
proportion of treated neighbors, a threshold indicator, or an
application-specific measure of social influence
\citep{aronow2017estimating,forastiere2021identification,
forastiere2022estimating,li2022random,ogburn2022causal}. In this study, we adopt the weighted
proportion of treated neighbors, which
maps neighborhoods of different sizes to \([0,1]\) and reduces to the
ordinary treated-neighbor proportion in an unweighted network:
\[
    \phi
    \left(
        \boldsymbol a_{k[\mathcal N_{ki}]};
        \boldsymbol w_{ki}
    \right)
    =
    \frac{
        \sum_{j\in\mathcal N_{ki}}w_{kij}a_{kj}
    }{
        \sum_{j\in\mathcal N_{ki}}w_{kij}
    }.
\]

We similarly summarize neighboring pretreatment covariates. Let
\(\boldsymbol X_{ki}\in\mathbb R^r\) denote the pretreatment
covariate vector of individual \(i\), and let
\(\boldsymbol X_{k[\mathcal N_{ki}]}
=(\boldsymbol X_{kj}^{\top})_{j\in\mathcal N_{ki}}
\in\mathbb R^{d_{ki}\times r}\)
collect the covariates of its neighbors.

\begin{definition}[Neighborhood-covariate aggregation]
\label{def:neighbor-covariates}

Let $\psi:
    \bigcup_{d\geq1}(
        \mathbb R^{d\times r}\times\mathbb R_{+}^{d}
    )
    \rightarrow
    \mathbb R^{r^{\mathrm{nbr}}}$
be a prespecified aggregation rule. For individual \(i\) in network
\(k\), define the fixed-dimensional
summary of the pretreatment covariates in the neighborhood as
\[
    \boldsymbol X_{ki}^{\mathrm{nbr}}
    =
    \psi
    \left(
        \boldsymbol X_{k[\mathcal N_{ki}]};
        \boldsymbol w_{ki}
    \right)
    \in\mathbb R^{r^{\mathrm{nbr}}}.
\]
\end{definition}

A common choice of \(\psi\) is also the weighted average of neighboring
covariates, under which \(r^{\mathrm{nbr}}=r\). We then combine the
individual covariates with their neighborhood summary and append the
neighborhood exposure to form the model input $\boldsymbol{U}_{ki}$. Specifically,
\begin{equation}
\begin{aligned}
    \boldsymbol X_{ki}^{\mathrm{nbr}}
    &=
    \frac{
        \sum_{j\in\mathcal N_{ki}}
        w_{kij}\boldsymbol X_{kj}
    }{
        \sum_{j\in\mathcal N_{ki}}w_{kij}
    },\\
    \boldsymbol Z_{ki}
    &=
    \begin{pmatrix}
        \boldsymbol X_{ki}\\
        \boldsymbol X_{ki}^{\mathrm{nbr}}
    \end{pmatrix}
    \in\mathbb R^h,
    \qquad
    h=r+r^{\mathrm{nbr}},\\
    \boldsymbol U_{ki}
    &=
    \begin{pmatrix}
        E_{ki}\\
        \boldsymbol Z_{ki}
    \end{pmatrix}
    \in\mathbb R^{h+1}.
\end{aligned}
\label{eq:augmented-input}
\end{equation}

For arbitrary intervention levels \(a\in\{0,1\}\) and
\(e\in\mathcal E\), let \(Y_{ki}(a,e)\) denote the potential outcome
of individual \(i\) in network \(k\) under own treatment \(a\) and
neighborhood exposure \(e\). The two aggregation mappings defined above extend the individual covariate
input to \(\boldsymbol u=(e,\boldsymbol z^\top)^\top\),
incorporating neighborhood exposure and aggregated neighboring covariates.
We then characterize individual heterogeneity through the
treatment-specific conditional response surface
\[
    f_a(\boldsymbol u)
    =
    f_a(e,\boldsymbol z)
    =
    \mathbb E
    \left[
        Y_{ki}(a,e)
        \mid
        \boldsymbol Z_{ki}=\boldsymbol z
    \right].
\]

The following assumptions establish identification of \(f_a(e,\boldsymbol z)\) at attainable inputs and its stability across networks. Conditions~(A1)--(A3) jointly identify the conditional causal
response surface at such configurations as
$f_a(e,\boldsymbol z)
    =
    \mathbb E
    \left[
        Y_{ki}
        \mid
        A_{ki}=a,\,
        E_{ki}=e,\,
        \boldsymbol Z_{ki}=\boldsymbol z
    \right]$
\citep{forastiere2021identification,forastiere2022estimating}.
Condition~(A4) ensures that the same response surface applies across
networks, thereby allowing observations from different networks to be
pooled.
\begin{assumption}[Identification and Stability]
\label{assumption:identification}
For every \(a\in\{0,1\}\), \(e\) in the attainable target exposure
support, and \(\boldsymbol z\) in the target covariate support, the following conditions hold:
\begin{enumerate}[label=\textnormal{(A\arabic*)}]
    \item \textbf{Consistency.} The observed outcome equals the
potential outcome: $Y_{ki}=Y_{ki}(A_{ki},E_{ki})$.

    \item \textbf{Exchangeability.}  The potential outcome is conditionally independent of own treatment and neighborhood exposure given the pretreatment covariates:
    \(Y_{ki}(a,e)\perp\!\!\!\perp
    (A_{ki},E_{ki})\mid\boldsymbol Z_{ki}\).

    \item \textbf{Positivity.}
    The conditional joint probability mass satisfies
\(\Pr(A_{ki}=a,E_{ki}=e\mid
\boldsymbol Z_{ki}=\boldsymbol z)>0\).

    \item \textbf{Stability.}
    There exists a common \(f_a\) such that
    \(\mathbb E[Y_{ki}(a,e)\mid\boldsymbol Z_{ki}=\boldsymbol z]
    =f_a(e,\boldsymbol z)\) for all \(k\in\mathcal K\).
\end{enumerate}
\end{assumption}

The identification in Assumption~\ref{assumption:identification} applies to configurations supported by the assignment mechanism; feasibility alone does not ensure positivity. With a finite network pool, attainable exposure levels may be discrete, so recovery between these levels additionally relies on the smoothness and coverage conditions used below.

\subsection{Causal Estimands and Sequential Experimental Design}
\label{sec:estimands-design-problem}

In this study, the causal estimands of interest are the population-average response
functions and the causal effect functions derived from them. Given the
common conditional response surface \(f_a(e,\boldsymbol z)\), we
average over the target covariate distribution \(P_{\boldsymbol Z}\)
to define the population-average response under own treatment \(a\)
and neighborhood exposure \(e\):
\begin{equation}
    m_a(e)
    =
    \mathbb E_{\boldsymbol Z}
    \left[
        f_a(e,\boldsymbol Z)
    \right]
    =
    \int
        f_a(e,\boldsymbol z)
        \,dP_{\boldsymbol Z}(\boldsymbol z).
\label{eq:population-response}
\end{equation}

Contrasts of the population-average responses define the direct,
spillover, and overall effect estimands \citep{hudgens2008toward,forastiere2021identification,forastiere2022estimating}. The direct effect varies own
treatment at a fixed exposure, the spillover effect varies exposure
under no own treatment, and the overall effect
combines both changes relative to the untreated and unexposed reference
condition. These contrasts satisfy
\(\tau_{\mathrm{overall}}(e)
=\tau_{\mathrm{dir}}(e)+\tau_{\mathrm{spill}}(e)\).

\begin{definition}[Causal effect functions]
\label{def:causal-effects}

The direct, spillover, and overall effect functions are
\begin{equation}
\begin{aligned}
    \tau_{\mathrm{dir}}(e)
    &=
    m_1(e)-m_0(e),\\
    \tau_{\mathrm{spill}}(e)
    &=
    m_0(e)-m_0(0),\\
    \tau_{\mathrm{overall}}(e)
    &=
    m_1(e)-m_0(0).
\end{aligned}
\label{eq:causal-effects}
\end{equation}
\end{definition}

To learn the population-average response functions and causal effect functions efficiently under a
limited experimental budget, we address three key questions:
(1) how to estimate the response surfaces \(f_0\) and \(f_1\), and
thereby the population-average responses and causal effect functions,
from the data collected thus far; (2) how to construct an acquisition
criterion \(\mathcal J_{t+1}\) for evaluating and selecting the next
batch of input locations; and (3) how to solve the resulting
experimental-design problem efficiently.

We now formalize the sequential experimental framework underlying both
estimation and design. For each candidate network \(k\), suppose that
the network matrix \(\boldsymbol W_k\) and the covariate matrix
\(\boldsymbol z_k
=(\boldsymbol z_{k1}^{\top},\ldots,
\boldsymbol z_{kn_k}^{\top})^\top
\in\mathbb R^{n_k\times h}\)
are known before treatment assignment. Outcomes, however, are observed
only after a network has been selected and its treatment assignment
implemented.  After completing round \(t\), the data collected during
the first \(t\) rounds are
\begin{equation*}
    \mathcal D_t
    =
    \left\{
        \left(
            a_{k_\ell i}^{(\ell)},
            \boldsymbol u_{k_\ell i}^{(\ell)},
            y_{k_\ell i}^{(\ell)}
        \right):
        \ell=1,\ldots,t,\ 
        i\in\mathcal V_{k_\ell}
    \right\},
\label{eq:observed-data}
\end{equation*}
where the implemented assignment
\(\boldsymbol a_{k_\ell}^{(\ell)}\) on network
\(k_\ell\) induces the exposure
\(e_{k_\ell i}^{(\ell)}
=\phi(\boldsymbol a_{k_\ell[\mathcal N_{k_\ell i}]}^{(\ell)};
\boldsymbol w_{k_\ell i})\), and
\(\boldsymbol u_{k_\ell i}^{(\ell)}
=(e_{k_\ell i}^{(\ell)},
\boldsymbol z_{k_\ell i}^{\top})^\top\)
is the corresponding exposure--covariate input, incorporating both
individual and neighborhood information.

Our objective is to select the next batch of input locations to improve
estimation of the population-average response functions and causal effect functions. In an unconstrained
problem with batch size \(n\), the input locations can be selected
freely from the target domain \(\Omega\) by minimizing the
prespecified criterion:
\begin{equation}
    \left(
        \boldsymbol u_i^{(t+1)}
    \right)_{i=1}^{n}
    \in
    \underset{\boldsymbol u_1,\ldots,\boldsymbol u_n\in\Omega}
    {\operatorname{argmin}}\,
    \mathcal J_{t+1}
    \left(
        \boldsymbol u_1,\ldots,\boldsymbol u_n
        \mid\mathcal D_t
    \right).
\label{eq:unconstrained-active-design}
\end{equation}

Under network interference, however, the input locations cannot be
selected independently, unlike in conventional batch sequential design
\citep{loeppky2010batch}. For a candidate network \(k\), the network
structure \(\boldsymbol W_k\) and assignment vector
\(\boldsymbol a_k\) jointly determine the input batch
\(\{\boldsymbol u_{ki}:i\in\mathcal V_k\}\). Thus, the assignment and
the resulting inputs are coupled through the network, distinguishing
the present problem from conventional batch sequential design.

%%%%%%%%%%%%%%%%%%%%%%%%%%%%%%%%%%%%%%%%%%%%%%%%%%%%%%%%%%%%%%%%%%%%%%
\section{Methodology}
\label{sec:methodology}

This section introduces the \FANCI \  framework.
Section~\ref{sec:gp-learning} develops GP-based estimation of the
population-average response functions and causal effect functions.
Section~\ref{sec:network-active-design} formulates a fill-distance
criterion for active learning under network interference, and
Section~\ref{sec:solution-strategies} presents full-network and
target-screened optimization strategies for selecting the next
experiment.

\subsection{Gaussian-Process Learning}
\label{sec:gp-learning}

Under Assumption~\ref{assumption:identification}, observations from
different networks identify the same treatment-specific conditional
response surface \(f_a(\boldsymbol u)\). We therefore pool
observations across completed network experiments for each
own-treatment level and fit \(f_0\) and \(f_1\) separately. 
For \(a\in\{0,1\}\), let \(n_a^{(t)}\) denote the number of
observations in \(\mathcal D_t\) with own treatment \(a\), and let
\(s=1,\ldots,n_a^{(t)}\) index these observations stacked across
experimental rounds in an arbitrary but fixed order.
We define the corresponding input--outcome pairs as
\(\{
    (
        \boldsymbol u_{a,s}^{(t)},
        y_{a,s}^{(t)}
    )
\}_{s=1}^{n_a^{(t)}}
=
\{
    (
        \boldsymbol u_{k_\ell i}^{(\ell)},
        y_{k_\ell i}^{(\ell)}
    ):
    \ell=1,\ldots,t,\ 
    i\in\mathcal V_{k_\ell},\
    a_{k_\ell i}^{(\ell)}=a
\}\). We assume that each treatment-specific conditional response surface follows a
GP:
\begin{equation}
\begin{aligned}
    f_a &\sim \operatorname{GP}(\mu_a,\kappa_a),\\
    y_{a,s}^{(t)}
    &=f_a\!\left(\boldsymbol u_{a,s}^{(t)}\right)
      +\varepsilon_{a,s},
\end{aligned}
\label{eq:gp-model}
\end{equation}
where
\(\varepsilon_{a,s}\overset{\mathrm{iid}}{\sim}
\mathcal N(0,\sigma_a^2)\).
We use this GP specification as a working model for estimation; the theoretical analysis treats the response surface as a fixed function satisfying Assumption~\ref{assumption:learning-regularity}.
The mean function \(\mu_a\), kernel \(\kappa_a\), and noise variance
\(\sigma_a^2\) may differ between the two treatment levels. By placing a GP model on the entire response surface, this formulation flexibly learns nonlinear conditional response surfaces without imposing a parametric form \citep{rasmussen2006gaussian,gramacy2020surrogates}. The kernel quantifies uncertainty across the exposure--covariate domain and provides information for subsequent experimental design.

In estimation, \(\mu_a\) may either be prespecified, such as a zero
mean after treatment-specific outcome centering, or parameterized and
estimated jointly with the covariance parameters. The kernel
\(\kappa_a\) can likewise be specified flexibly. For example, an
anisotropic kernel may assign different length scales to the
neighborhood exposure \(e\) and the individual components of
\(\boldsymbol z\). Parameters in the mean function, together with the
kernel hyperparameters and noise variance, can be estimated by
maximizing the treatment-specific marginal likelihood
\citep{rasmussen2006gaussian}. Appendix~A
provides a concrete kernel specification and details the corresponding
estimation procedure. When the accumulated sample becomes too large
for exact GP inference, regularized, sparse, or otherwise scalable GP
approximations may be used
\citep{li2026gpbss,chokhachian2026fastgp}.

After fitting the GP models using \(\mathcal D_t\), we obtain the
posterior distribution of the conditional response surface at any
input \(\boldsymbol u\). Let
\(\boldsymbol y_a^{(t)}
=(y_{a,1}^{(t)},\ldots,y_{a,n_a^{(t)}}^{(t)})^\top\)
collect the outcomes with own treatment \(a\), and define
\(\boldsymbol\mu_a^{(t)}
=(\mu_a(\boldsymbol u_{a,1}^{(t)}),\ldots,
\mu_a(\boldsymbol u_{a,n_a^{(t)}}^{(t)}))^\top\)
as the corresponding prior mean vector. Define
\([\boldsymbol K_a^{(t)}]_{ss'}=
\kappa_a(\boldsymbol u_{a,s}^{(t)},\boldsymbol u_{a,s'}^{(t)})\)
and
\([\boldsymbol k_a^{(t)}(\boldsymbol u)]_s=
\kappa_a(\boldsymbol u_{a,s}^{(t)},\boldsymbol u)\).
For compactness, set
\(\boldsymbol Q_a^{(t)}
=(\boldsymbol K_a^{(t)}+\sigma_a^2\boldsymbol I)^{-1}\).
Conditional on \(\mathcal D_t\), the posterior mean and covariance of
the latent response surface are
\begin{equation}
\begin{aligned}
    \widehat f_a^{(t)}(\boldsymbol u)
    ={}&
    \mu_a(\boldsymbol u)
    +
    \boldsymbol k_a^{(t)}(\boldsymbol u)^\top
    \boldsymbol Q_a^{(t)}
    \left(
        \boldsymbol y_a^{(t)}
        -
        \boldsymbol\mu_a^{(t)}
    \right),\\
    \widehat\kappa_a^{(t)}
    (\boldsymbol u,\boldsymbol u')
    ={}&
    \kappa_a(\boldsymbol u,\boldsymbol u')
    -
    \boldsymbol k_a^{(t)}(\boldsymbol u)^\top
    \boldsymbol Q_a^{(t)}
    \boldsymbol k_a^{(t)}(\boldsymbol u').
\end{aligned}
\label{eq:gp-posterior}
\end{equation}
%The posterior mean provides a point estimate of the conditional causal response at both observed and unobserved inputs, whereas the posterior covariance characterizes the remaining joint uncertainty across different inputs.

We next obtain the population-average response in
Equation~\eqref{eq:population-response} by standardizing the fitted
conditional surface over the target covariate distribution, which is
assumed to be available before experimentation. Recall
from Equation~\eqref{eq:augmented-input} that
\(\boldsymbol u=(e,\boldsymbol z^\top)^\top\in\mathbb R^{h+1}\).
Because population standardization is a linear operator, the induced
posterior process for \(m_a\) remains Gaussian. Its posterior mean and
covariance are
\begin{equation}
\small
\begin{aligned}
    \widehat m_a^{(t)}(e)
    ={}&
    \frac{1}{n_{\mathrm{pop}}}
    \sum_{q=1}^{n_{\mathrm{pop}}}
    \widehat f_a^{(t)}
    \left(
        (e,\boldsymbol z_q^{\mathrm{pop}\top})^\top
    \right),\\
    \widehat v_a^{(t)}(e,e')
    ={}&
    \frac{1}{n_{\mathrm{pop}}^2}
    \sum_{q=1}^{n_{\mathrm{pop}}}
    \sum_{q'=1}^{n_{\mathrm{pop}}}
    \widehat\kappa_a^{(t)}
    \left(
        (e,\boldsymbol z_q^{\mathrm{pop}\top})^\top,
        (e',\boldsymbol z_{q'}^{\mathrm{pop}\top})^\top
    \right),
\end{aligned}
\label{eq:population-posterior}
\end{equation}
where
\(\{\boldsymbol z_q^{\mathrm{pop}}\}_{q=1}^{n_{\mathrm{pop}}}\)
denotes a covariate sample representing the target population
distribution \(P_{\boldsymbol Z}\).
The reference sample may comprise the full deployment population or a
representative sample drawn from the target population. When
expectations with respect to \(P_{\boldsymbol Z}\) are analytically
tractable, the empirical averages in
Equation~\eqref{eq:population-posterior} may be replaced by their
corresponding population integrals.

Substituting the standardized responses into
Definition~\ref{def:causal-effects} gives
\begin{equation}
\begin{aligned}
\widehat\tau_{\mathrm{dir}}^{(t)}(e)
&=\widehat m_1^{(t)}(e)
 -\widehat m_0^{(t)}(e),\\
\widehat\tau_{\mathrm{spill}}^{(t)}(e)
&=\widehat m_0^{(t)}(e)
 -\widehat m_0^{(t)}(0),\\
\widehat\tau_{\mathrm{overall}}^{(t)}(e)
&=\widehat m_1^{(t)}(e)
 -\widehat m_0^{(t)}(0).
\end{aligned}
\label{eq:effect-estimators}
\end{equation}

\subsection{Active Learning under Network Interference}
\label{sec:network-active-design}

Given the data collected through round \(t\), the active-learning
procedure selects the experiment for round \(t+1\) and thereby
determines the input locations it will generate. As discussed in
Section~\ref{sec:estimands-design-problem}, these locations cannot be
selected independently under network interference. For a candidate
network \(k\), its topology \(\boldsymbol W_k\) and covariate profiles
\(\boldsymbol z_k\) are fixed, while a single binary assignment vector
\(\boldsymbol a_k\) jointly determines the own-treatment levels and
neighborhood exposures of all individuals. Thus, rather than choosing
the input locations directly, the experimenter selects a feasible network--assignment pair
\((k,\boldsymbol a_k)\). Accordingly, the unconstrained optimization
over freely chosen inputs in
Equation~\eqref{eq:unconstrained-active-design} is replaced by an
optimization over such feasible pairs.

Let
\(\mathcal U_a^{(t)}
=\{\boldsymbol u_{a,s}^{(t)}:
s=1,\ldots,n_a^{(t)}\}\)
denote the input locations observed through round \(t\) for treatment
level \(a\). For a candidate network--assignment pair
\((k,\boldsymbol a_k)\), let
\(\boldsymbol u_{ki}\) denote the input induced for individual
\(i\in\mathcal V_k\). The augmented design set resulting from
implementing this pair is
\begin{equation*}
    \mathcal U_a^{(t+1)}(k,\boldsymbol a_k)
    =
    \mathcal U_a^{(t)}
    \cup
    \left\{
        \boldsymbol u_{ki}:
        i\in\mathcal V_k,\ a_{ki}=a
    \right\}.
\label{eq:candidate-design-set}
\end{equation*}

We aim to select \((k,\boldsymbol a_k)\) such that the
augmented design sets
\(\mathcal U_a^{(t+1)}(k,\boldsymbol a_k)\),
\(a\in\{0,1\}\), are as space-filling as possible over their
respective target domains. Because the goal is to recover population-average response functions
and causal effect functions globally over these domains, the design should
improve global coverage rather than concentrate observations in a local
region. Fill distance provides
a natural criterion for this purpose: it measures the greatest distance
from any target input to its nearest design point and is closely related
to uniform response-surface approximation
\citep{wendland2004scattered,loeppky2010batch,kanagawa2018gaussian}. For each
\(a\in\{0,1\}\), the fill distance associated with candidate pair
\((k,\boldsymbol a_k)\) is
\begin{equation}
    \FD_{a}^{(t+1)}(k,\boldsymbol a_k)
    =
    \sup_{\boldsymbol v\in\Omega_a}
    \inf_{\boldsymbol u\in
    \mathcal U_a^{(t+1)}(k,\boldsymbol a_k)}
    \|\boldsymbol v-\boldsymbol u\|_{\boldsymbol M_a},
\label{eq:joint-fill-distance}
\end{equation}
where \(\Omega_a\subseteq\mathbb R^{h+1}\) is a compact target domain
covering the exposure range of interest and the relevant support of
the target covariates. For a
positive definite weighting matrix \(\boldsymbol M_a\),
\(\|\boldsymbol u\|_{\boldsymbol M_a}
=(\boldsymbol u^\top\boldsymbol M_a\boldsymbol u)^{1/2}\)
denotes the corresponding weighted Euclidean norm. 

We combine the two treatment-specific fill distances as
\begin{equation}
    \mathcal J_{t+1}^{\mathrm{FD}}(k,\boldsymbol a_k)
    =
    \sum_{a\in\{0,1\}}
    \omega_a\,
    \FD_a^{(t+1)}(k,\boldsymbol a_k).
\label{eq:fill-distance-acquisition}
\end{equation}
where \(\omega_0,\omega_1\geq0\) and
\(\omega_0+\omega_1=1\). Equal weights place symmetric emphasis on
the two response surfaces, whereas unequal weights may prioritize an
underrepresented or substantively more important treatment level.
Both weights should remain positive when the direct and overall effect
functions are primary inferential targets.

A smaller fill
distance places a tighter upper bound on the distance between any
target input and its nearest point in the augmented design set, thereby
indicating better worst-case coverage of \(\Omega_a\). Consequently, the network and assignment for round \(t+1\) are
selected by solving
\begin{equation}
\left(
    k_{t+1},
    \boldsymbol a_{k_{t+1}}^{(t+1)}
\right)
\in
\underset{
    k\in\mathcal K_{t+1},\,
    \boldsymbol a_k\in\mathcal A_k
}{
    \operatorname{argmin}
}
\mathcal J_{t+1}^{\mathrm{FD}}
(k,\boldsymbol a_k),
\label{eq:fill-distance-design}
\end{equation}
where
\(\mathcal K_{t+1}\subseteq\mathcal K\) denote the networks available
after round \(t\). Under the assumption that each network can be
selected at most once, \(\mathcal K_1=\mathcal K\) and
\(\mathcal K_{t+1}
=\mathcal K_t\setminus\{k_t\}\) for \(t\geq1\). In addition,
\(\mathcal A_k\subseteq\{0,1\}^{n_k}\) denote the set of feasible
treatment assignments, with
\(\mathcal A_k=\{0,1\}^{n_k}\) when no additional restrictions are
imposed. In practice, \(\mathcal A_k\) may encode treatment budgets,
bounds on treatment-group sizes, individual eligibility restrictions,
capacity limits, fairness requirements, or other operational
constraints.

Even without additional assignment constraints, a network with
\(n_k\) individuals may admit as many as \(2^{n_k}\) binary
assignments. Moreover, changing a single assignment component can
alter the individual's own-treatment level and the neighborhood
exposures of multiple connected individuals, thereby changing both
the induced inputs and their treatment-specific design-set membership.
The resulting objective is generally nonconvex, nonsmooth, and
nonseparable across individuals. Hence,
Equation~\eqref{eq:fill-distance-design} defines a large-scale
combinatorial optimization problem.

\subsection{Full-Network and Target-Screened Optimization}
\label{sec:solution-strategies}

In this section, we develop two computational strategies for solving the combinatorial design problem: full-network optimization and target-screened optimization. The former emphasizes solution quality, whereas the latter reduces computation through candidate screening.

\paragraph{Full-network optimization.}
The full-network strategy evaluates each candidate experiment by the
global coverage of its induced input set. Since
Equation~\eqref{eq:fill-distance-design} can be decomposed across
networks without changing the objective, for each
\(k\in\mathcal K_{t+1}\), we first solve
\begin{equation}
    \boldsymbol a_k^{\star(t+1)}
    \in
    \underset{\boldsymbol a_k\in\mathcal A_k}
    {\operatorname{argmin}}\,
    \mathcal J_{t+1}^{\mathrm{FD}}
    (k,\boldsymbol a_k).
\label{eq:within-network-assignment}
\end{equation}

We then select
\begin{equation}
    k_{t+1}
    \in
    \underset{k\in\mathcal K_{t+1}}
    {\operatorname{argmin}}\,
    \mathcal J_{t+1}^{\mathrm{FD}}
    \bigl(k,\boldsymbol a_k^{\star(t+1)}\bigr).
\label{eq:network-selection}
\end{equation}

Solving Equations~\eqref{eq:within-network-assignment} and
\eqref{eq:network-selection} exactly is equivalent to direct
optimization over \(\mathcal J_{t+1}\), and the within-network
problems can be solved in parallel. For small networks, the within-network problem can be solved by
enumeration or an exact discrete optimizer. For larger networks, we
use a genetic algorithm \citep{goldberg1989genetic} to search over
binary treatment assignments
\(\boldsymbol a_k\in\mathcal A_k\).
Each assignment induces the corresponding exposures and inputs for
evaluating Equation~\eqref{eq:fill-distance-acquisition}.
Additional assignment constraints can be handled through feasible
initialization and repair operators.

\paragraph{Target-screened optimization.}
When many large networks are available, solving the assignment problem
for every candidate network can be costly. The target-screened strategy
first identifies informative input regions, uses them to screen the
available networks, and then solves the original objective over a reduced candidate set. For each treatment level \(a\in\{0,1\}\), we
consider two target constructions:
\begin{enumerate}[label=\textnormal{(\roman*)}]
    \item \textit{Coverage-guided target.}
    To target the largest remaining geometric gap, we select
    \begin{equation}
        \boldsymbol u_{a,t+1}^{\dagger}
        \in
        \underset{\boldsymbol u\in\Omega_a}
        {\operatorname{argmax}}\,
        \inf_{\boldsymbol v\in\mathcal U_a^{(t)}}
        \|\boldsymbol u-\boldsymbol v\|_{\boldsymbol M_a}.
    \label{eq:coverage-screening-target}
    \end{equation}

    \item \textit{Uncertainty-guided target.}
    To target the region with the largest GP uncertainty \citep{schulz2018tutorial,chang2021data}, we select
    
    \begin{equation}
        \boldsymbol u_{a,t+1}^{\dagger}
        \in
        \underset{\boldsymbol u\in\Omega_a}
        {\operatorname{argmax}}\,
        \widehat\kappa_a^{(t)}
        (\boldsymbol u,\boldsymbol u).
    \label{eq:uncertainty-screening-target}
    \end{equation}
\end{enumerate}

The targets
$\boldsymbol u_{0,t+1}^{\dagger}$ and
$\boldsymbol u_{1,t+1}^{\dagger}$
identify regions where additional information is most valuable.
For each candidate network $k\in\mathcal K_{t+1}$,
we sample a few feasible assignments and compute a screening score
$\mathcal S_{t+1}(k)$, defined as the minimum weighted sum of
the distances from the two targets to their nearest
treatment-specific inputs across the sampled assignments.
Networks with smaller scores can better approximate the desired
targets. We retain the $R$ networks with the smallest scores,
yielding a reduced candidate set
$\widetilde{\mathcal K}_{t+1}\subseteq\mathcal K_{t+1}$.
We then solve the original fill-distance design problem only
over this screened set:
\begin{equation}
\left(
    k_{t+1},
    \boldsymbol a_{k_{t+1}}^{(t+1)}
\right)
\in
\underset{
    k\in\widetilde{\mathcal K}_{t+1},\,
    \boldsymbol a_k\in\mathcal A_k
}{
    \operatorname{argmin}
}
\mathcal J_{t+1}^{\mathrm{FD}}(k,\boldsymbol a_k).
\label{eq:screened-fill-distance-design}
\end{equation}

This screening step reduces the number of expensive within-network
assignment problems from \(|\mathcal K_{t+1}|\) to \(R\), although the
problem within each retained network remains combinatorial. Its
effectiveness therefore depends on retaining networks capable of
generating high-quality designs. The parameter \(R\) controls the
trade-off between computational cost and screening breadth, with
\(R=|\mathcal K_{t+1}|\) recovering full-network optimization.

After either strategy selects
\((k_{t+1},\boldsymbol a_{k_{t+1}}^{(t+1)})\), the experiment is
implemented, yielding
\(\mathcal D_{t+1}^{\mathrm{new}}
=\{(a_{k_{t+1}i}^{(t+1)},
\boldsymbol u_{k_{t+1}i}^{(t+1)},
y_{k_{t+1}i}^{(t+1)}):
i\in\mathcal V_{k_{t+1}}\}\).
We update
\(\mathcal D_{t+1}
=\mathcal D_t\cup\mathcal D_{t+1}^{\mathrm{new}}\) and
\(\mathcal K_{t+2}
=\mathcal K_{t+1}\setminus\{k_{t+1}\}\), refit the
treatment-specific GP response surfaces, and repeat until the
experimental budget is exhausted or a prespecified stopping criterion
is met. Algorithm~\ref{alg:fanci} summarizes the complete FANCI framework.

\begin{algorithm}[!h]
\caption{\FANCI\ for Sequential Experiments under Network Interference}
\label{alg:fanci}
\begin{algorithmic}[1]
\STATE \textbf{Input:} Initial data \(\mathcal D_0\), candidate
networks \(\mathcal K_1\), feasible assignment sets
\(\{\mathcal A_k\}\), target domains \(\Omega_0,\Omega_1\), weights
\(\omega_0,\omega_1\), screening size \(R\), solution strategy, and
experimental budget \(T\).

\FOR{\(t=0,\ldots,T-1\)}
    \STATE Fit the treatment-specific GP models using
    \(\mathcal D_t\) and obtain
    \(\widehat f_a^{(t)}\) and
    \(\widehat\kappa_a^{(t)}\) from
    Equation~\eqref{eq:gp-posterior}.

    \STATE Standardize the fitted response surfaces to obtain
    \(\widehat m_a^{(t)}\) and update the causal effect functions using
    Equations~\eqref{eq:population-posterior} and
    \eqref{eq:effect-estimators}.

   \STATE Under full-network optimization, select
\((k_{t+1},\boldsymbol a_{k_{t+1}}^{(t+1)})\) using
Equations~\eqref{eq:within-network-assignment} and
\eqref{eq:network-selection}. Under target-screened optimization,
identify the targets using Equation~\eqref{eq:coverage-screening-target}
or \eqref{eq:uncertainty-screening-target}, and select
\((k_{t+1},\boldsymbol a_{k_{t+1}}^{(t+1)})\) using
Equation
\eqref{eq:screened-fill-distance-design}.

    \STATE Implement the selected experiment, observe
    \(\mathcal D_{t+1}^{\mathrm{new}}\), and update
    \(\mathcal D_{t+1}\) and \(\mathcal K_{t+2}\).
\ENDFOR

\STATE \textbf{Output:} Conditional response surfaces and population-average response functions
\(\{\widehat f_a^{(T)},\widehat m_a^{(T)}:a\in\{0,1\}\}\) and effect
functions \(\widehat\tau_{\mathrm{dir}}^{(T)}\),
\(\widehat\tau_{\mathrm{spill}}^{(T)}\), and
\(\widehat\tau_{\mathrm{overall}}^{(T)}\).
\end{algorithmic}
\end{algorithm}

%%%%%%%%%%%%%%%%%%%%%%%%%%%%%%%%%%%%%%%%%%%%%%%%%%%%%%%%%%%%%%%%%%%%%%
%%%%%%%%%%%%%%%%%%%%%%%%%%%%%%%%%%%%%%%%%%%%%%%%%%%%%%%%%%%%%%%%%%%%%
\section{Theoretical Guarantees}
\label{sec:theory}

This section establishes two theoretical properties of \FANCI:
statistical error control under fill-distance-guided learning and an
approximation guarantee for target-screened optimization. All proofs
are provided in Appendix~B.

We first analyze the accuracy of the estimated response and causal
effect functions. Let \(\mathcal H_{\kappa_a}\) denote the reproducing
kernel Hilbert space (RKHS) induced by \(\kappa_a\). We use
\(\|\cdot\|_{\infty,\Omega_a}\) and
\(\|\cdot\|_{\infty,\mathcal E}\) to denote the supremum norms over
\(\Omega_a\) and \(\mathcal E\), respectively. We impose the following
regularity conditions.

\begin{assumption}[Sequential GP Regularity]
\label{assumption:learning-regularity}
For each \(a\in\{0,1\}\), the following conditions hold:
\begin{enumerate}[
label=\textnormal{(A\arabic*)},
start=5
]

\item \textbf{Target support.}
The compact domain \(\Omega_a\) contains all exposure--covariate inputs
required for population standardization. For finite network pools,
\(\Omega_a\) need not be restricted to the discrete set of attainable inputs.

\item \textbf{Kernel regularity.}
For the fixed mean function \(\mu_a\), the centered response surface satisfies
\(f_a-\mu_a\in\mathcal H_{\kappa_a}\) with
\(\|f_a-\mu_a\|_{\mathcal H_{\kappa_a}}\le B_a<\infty\).
The kernel admits the representation
$\kappa_a(\boldsymbol u,\boldsymbol v)
=
\sigma_{\kappa,a}^{2}
R_a\!\left(
d_{\kappa,a}(\boldsymbol u,\boldsymbol v)
\right),$
where \(R_a:[0,\infty)\to[0,1]\) is continuous and nonincreasing
with \(R_a(0)=1\),
and, for some \(c_a<\infty\), $d_{\kappa,a}(\boldsymbol u,\boldsymbol v)
\le
c_a
\|\boldsymbol u-\boldsymbol v\|_{\boldsymbol M_a}.$

\item \textbf{Stochastic estimation.}
Each observation error is conditionally mean-zero and \(\sigma_a\)-sub-Gaussian given the experimental history available before that outcome is observed. The GP mean, kernel, and noise variance are fixed in the theoretical analysis, with \(\sigma_a^2>0\). An explicit remainder \(r_{a,t}(\delta)\) controls the error relative to the noise-free kernel interpolant jointly for both treatment levels with probability at least \(1-\delta\).

\item \textbf{Population standardization.}
The expectation in Equation~\eqref{eq:population-posterior} is
evaluated exactly under \(P_{\boldsymbol Z}\), either analytically
or over the complete finite target population.

\end{enumerate}
\end{assumption}

Assumption~\ref{assumption:learning-regularity} separates the uniform
response-surface error into a geometric component governed by
experimental coverage and a stochastic component due to noisy
observations. Condition~(A5) ensures that the learning domain covers
the inputs relevant for population inference, while Condition~(A6)
aligns the kernel geometry with the distance used in the acquisition
criterion, allowing smaller fill distance to imply smaller uniform
approximation error. Condition~(A7) controls the remaining estimation error, including noise and regularization, and Condition~(A8) ensures that population
standardization introduces no additional approximation error. Appendix~B.1
discusses Conditions~(A1)--(A8), verifies the kernel-distance
condition in (A6), and gives a stochastic bound for (A7). For
common stationary kernels, sharper rates can be obtained under
standard domain and design regularity, including polynomial rates for
Mat\'ern kernels
\citep{wendland2004scattered,kanagawa2018gaussian}. The following
theorem propagates these bounds to the population-average response and
causal effect functions.

\begin{theorem}[Fill-Distance Error Bounds]
\label{thm:causal-effect-error}
Suppose that Assumptions~\ref{assumption:identification} and
\ref{assumption:learning-regularity} hold. For
\(a\in\{0,1\}\), define
\begin{equation*}
\begin{aligned}
\rho_a(h)
&=
B_a\sigma_{\kappa,a}
\left\{
1-R_a^2(c_a h)
\right\}^{1/2},\\
\varepsilon_{a,t}(\delta)
&=
\rho_a\!\left(\FD_a^{(t)}\right)
+
r_{a,t}(\delta).
\end{aligned}
\end{equation*}
Then \(\rho_a(h)\to0\) as \(h\downarrow0\). For any
\(\delta\in(0,1)\), with probability at least \(1-\delta\),
simultaneously for \(a\in\{0,1\}\),
\begin{equation*}
\begin{aligned}
\left\|
\widehat f_a^{(t)}-f_a
\right\|_{\infty,\Omega_a}
&\le
\varepsilon_{a,t}(\delta),\\
\left\|
\widehat m_a^{(t)}-m_a
\right\|_{\infty,\mathcal E}
&\le
\varepsilon_{a,t}(\delta).
\end{aligned}
\end{equation*}
Moreover,
\begin{equation*}
\begin{aligned}
\left\|
\widehat\tau_{\mathrm{dir}}^{(t)}
-
\tau_{\mathrm{dir}}
\right\|_{\infty,\mathcal E}
&\le
\varepsilon_{1,t}(\delta)
+
\varepsilon_{0,t}(\delta),\\
\left\|
\widehat\tau_{\mathrm{spill}}^{(t)}
-
\tau_{\mathrm{spill}}
\right\|_{\infty,\mathcal E}
&\le
2\varepsilon_{0,t}(\delta),\\
\left\|
\widehat\tau_{\mathrm{overall}}^{(t)}
-
\tau_{\mathrm{overall}}
\right\|_{\infty,\mathcal E}
&\le
\varepsilon_{1,t}(\delta)
+
\varepsilon_{0,t}(\delta).
\end{aligned}
\end{equation*}
\end{theorem}

Theorem~\ref{thm:causal-effect-error} yields uniform consistency when
both the geometric and stochastic errors vanish. Specifically, as the
treatment-specific designs become increasingly dense over
\(\Omega_0\) and \(\Omega_1\) and sufficient information accumulates
to attenuate the stochastic estimation error, the estimated response
and causal effect functions converge uniformly to their respective
targets. Because a fixed candidate pool permits only finitely many
network experiments, the asymptotic result is stated along an
expanding sequence of experimental problems. The following corollary
is conditional on increasing coverage and vanishing estimation error. 

\begin{corollary}[Uniform Consistency]
\label{cor:uniform-consistency}
Under the conditions of Theorem~\ref{thm:causal-effect-error},
suppose that, for every fixed \(\delta\in(0,1)\) and
\(a\in\{0,1\}\), \(\FD_a^{(t)}\to 0\) and
\(r_{a,t}(\delta)\to 0\) in probability. Then
\begin{equation*}
\begin{aligned}
\left\|
\widehat f_a^{(t)}-f_a
\right\|_{\infty,\Omega_a}
&\xrightarrow{p}0,\\
\left\|
\widehat m_a^{(t)}-m_a
\right\|_{\infty,\mathcal E}
&\xrightarrow{p}0,
\qquad a\in\{0,1\}.
\end{aligned}
\end{equation*}
In addition,
\begin{equation*}
\begin{aligned}
\left\|
\widehat\tau_{\mathrm{dir}}^{(t)}
-
\tau_{\mathrm{dir}}
\right\|_{\infty,\mathcal E}
&\xrightarrow{p}0,\\
\left\|
\widehat\tau_{\mathrm{spill}}^{(t)}
-
\tau_{\mathrm{spill}}
\right\|_{\infty,\mathcal E}
&\xrightarrow{p}0,\\
\left\|
\widehat\tau_{\mathrm{overall}}^{(t)}
-
\tau_{\mathrm{overall}}
\right\|_{\infty,\mathcal E}
&\xrightarrow{p}0.
\end{aligned}
\end{equation*}
\end{corollary}

We finally characterize the approximation loss introduced by target
screening. For each \(k\in\mathcal K_{t+1}\), let
\(\mathcal J_{t+1}^{\star}(k)\) denote the best fill-distance objective
attainable within network \(k\), and let
\(\mathcal J_{t+1}^{\star}\) and
\(\widetilde{\mathcal J}_{t+1}^{\star}\) denote the corresponding
optimal values before and after screening. In addition, let
\(\mathcal J_{t+1}^{\dagger}\) denote the weighted fill-distance
objective obtained by augmenting each treatment-specific design set
with its ideal target. Proposition~\ref{prop:screening-guarantee} links the target mismatch
measured by the screening score directly to the resulting fill-distance
objective. It shows that the screening
loss is controlled by the quality of the ideal-target design and by
how closely an available network can realize those targets. Thus,
target screening remains close to full-network optimization when the
ideal targets substantially improve coverage and can be closely
approximated by a feasible network experiment. The bound concerns exact within-network optimization; any additional error from a heuristic solver is separate.

\begin{proposition}[Target-Screening Approximation Bound]
\label{prop:screening-guarantee}
For every \(k\in\mathcal K_{t+1}\),
\begin{equation*}
\mathcal J_{t+1}^{\star}(k)
\le
\mathcal J_{t+1}^{\dagger}
+
\mathcal S_{t+1}(k).
\end{equation*}
If \(\widetilde{\mathcal K}_{t+1}\) contains the \(R\ge1\) networks
with the smallest screening scores, then
\begin{equation*}
\begin{aligned}
0
&\le
\widetilde{\mathcal J}_{t+1}^{\star}
-
\mathcal J_{t+1}^{\star}\\
&\le
\mathcal J_{t+1}^{\dagger}
+
\min_{k\in\mathcal K_{t+1}}
\mathcal S_{t+1}(k)
-
\mathcal J_{t+1}^{\star}.
\end{aligned}
\end{equation*}
\end{proposition}

\section{Numerical Experiments}
\label{sec:numerical}

This section evaluates \FANCI\ through controlled simulations and an
online-game case study. The numerical analysis compares alternative
network-constrained acquisition strategies, examines their accuracy in
learning the causal effect functions, and quantifies the consequences
of ignoring network interference.

\subsection{Experimental Setup}
\label{sec:simulation-setup}

\paragraph{Data generation.}

For each replication, we generate \(K=100\) candidate
networks, each containing \(n_k=100\) individuals. The adjacency
matrices \(\boldsymbol W_k\) are independently generated from an
Erd\H{o}s--R\'enyi random-graph model \citep{erdos1959random}. Any realization containing
isolated nodes is regenerated. 
Individual \(i\) in network \(k\) has three pretreatment covariates,
\(\boldsymbol X_{ki}=(X_{ki1},X_{ki2},X_{ki3})^\top\), where
\(X_{ki1}\) is binary and \(X_{ki2}\) and \(X_{ki3}\) are continuously
distributed on \([0,1]\). We construct \(E_{ki}\),
\(\boldsymbol X_{ki}^{\mathrm{nbr}}\in\mathbb R^3\),
\(\boldsymbol Z_{ki}\in\mathbb R^6\), and
\(\boldsymbol U_{ki}\in\mathbb R^7\) using the neighborhood-covariate
aggregation and exposure mapping defined in
Section~\ref{sec:causal-framework-interference}.

The potential outcome under own treatment \(a\in\{0,1\}\) and
neighborhood exposure \(e\in[0,1]\) is generated as
\begin{equation}
\begin{aligned}
Y_{ki}&(a,e)
=
g_a^{\mathrm{own}}(\boldsymbol X_{ki}) + (e-0.5) \cdot\\
&
\Bigl[
e\,g_1^{\mathrm{nbr}}(\boldsymbol X_{ki}^{\mathrm{nbr}})
+
(1-e)\,
g_0^{\mathrm{nbr}}(\boldsymbol X_{ki}^{\mathrm{nbr}})
\Bigr]
+\varepsilon_{ki},
\end{aligned}
\label{eq:simulation-outcome}
\end{equation}
where
\(g_a^{\mathrm{own}}(\boldsymbol x)
=\beta_{a1}x_1^2+\beta_{a2}x_2+\beta_{a3}/(1+x_3)\) and
\(g_a^{\mathrm{nbr}}(\boldsymbol x)
=\beta_{a1}^{\mathrm{nbr}}x_1^2
+\beta_{a2}^{\mathrm{nbr}}/(1+x_2)
+\beta_{a3}^{\mathrm{nbr}}x_3\).
The noise satisfies
\(\varepsilon_{ki}\sim\mathcal N(0,\sigma_\varepsilon^2)\), with the
coefficients and noise variance held fixed across replications. This
data-generating mechanism induces nonlinear and heterogeneous
dependence on both individual and neighborhood covariates, together
with nontrivial exposure--covariate interactions.

\paragraph{Competing methods.}
We compare three implementations of the proposed framework.
\FANCIFull\ evaluates the complete batch induced by each feasible
network--assignment pair under the fill-distance objective in
Equation~\eqref{eq:fill-distance-acquisition}. \FANCIGeo\ first screens
candidate networks using the coverage-guided targets in
Equation~\eqref{eq:coverage-screening-target}, whereas \FANCIUnc\ uses
the GP posterior-variance targets in
Equation~\eqref{eq:uncertainty-screening-target}; both then optimize
the original fill-distance objective over the retained networks. In
addition, we consider two benchmark methods. \MMC\ adapts the
maximum-model-change method of \citet{zhao2023batch} to the
network-constrained design space and selects the induced batch with
the largest information gain. \RAND\ is a nonadaptive random
network-and-assignment design: at each round, it selects an available
network uniformly at random, draws a common treatment probability
\(\pi_t\sim\operatorname{Unif}(0,1)\), and independently assigns
\(A_{ki}\sim\operatorname{Bernoulli}(\pi_t)\) within the selected
network.

\paragraph{Implementation details and evaluation criteria.}
For initialization, each replication begins with one common network
experiment whose treatment assignment generates observations near
\((a,e)=(0,0)\), \((1,1)\), \((0,0.5)\), and \((1,0.5)\). These
observations initialize both treatment-specific GP response surfaces
while leaving substantial portions of the exposure domain sparsely
covered. Starting from the second round, each method selects one
additional network experiment at each round and updates the estimated
response and causal effect functions after observing the resulting
outcomes.
The target population in
Equation~\eqref{eq:population-posterior} comprises the covariate
profiles, available before experimentation, of all \(10{,}000\)
individuals in the candidate-network pool.

We evaluate recovery of the overall and spillover effect functions in
Definition~\ref{def:causal-effects}. The direct effect is omitted from
the main comparison because
\(\tau_{\mathrm{dir}}(e)
=\tau_{\mathrm{overall}}(e)-\tau_{\mathrm{spill}}(e)\)
and is therefore fully determined by the two reported functions. For
replication \(p\), after \(t\) network experiments, and for
\(\tau\in\{\tau_{\mathrm{spill}},\tau_{\mathrm{overall}}\}\), define
\begin{equation}
\operatorname{ISE}^{(p)}(\tau;t)
=
\int_0^1
\left\{
\widehat\tau^{(p,t)}(e)-\tau^{(p)}(e)
\right\}^2\,de .
\label{eq:simulation-ise}
\end{equation}

The primary performance measure is the Monte Carlo mean integrated
squared error,
\(\operatorname{MISE}(\tau;t)
=100^{-1}\sum_{p=1}^{100}\operatorname{ISE}^{(p)}(\tau;t)\).
We repeat the complete experiment \(100\) times, independently
regenerating the networks, covariates, outcomes, and algorithmic
randomness. To ensure a controlled comparison, all methods use the same GP kernel family,
hyperparameter-estimation rule, target population, and
numerical-integration grid.

\subsection{Comparative Performance}
\label{sec:simulation-comparison}

\begin{figure}[p]
\centering

% FANCI-Full
\subfloat[\FANCIFull: 3 Networks\label{fig:full-3}]{
    \includegraphics[width=0.30\linewidth]
    {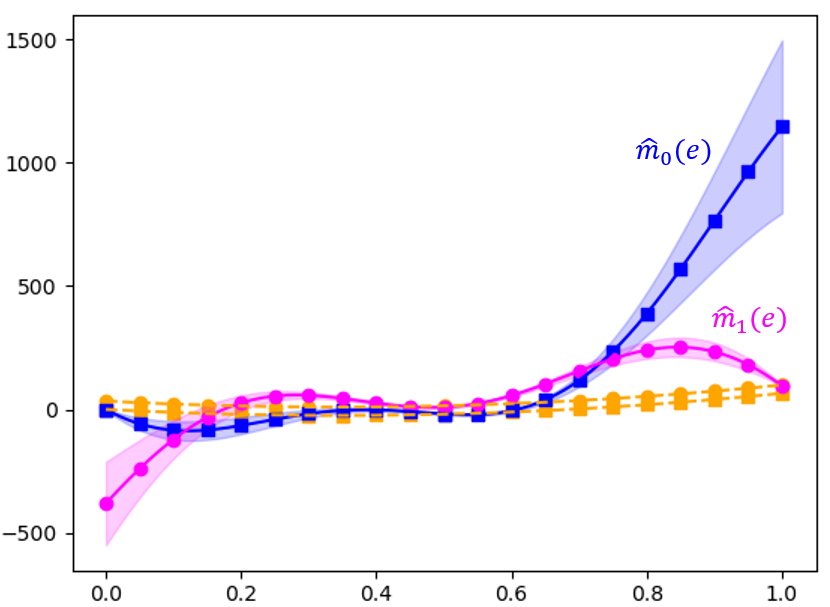}
}
\hfill
\subfloat[\FANCIFull: 5 Networks\label{fig:full-5}]{
    \includegraphics[width=0.30\linewidth]
    {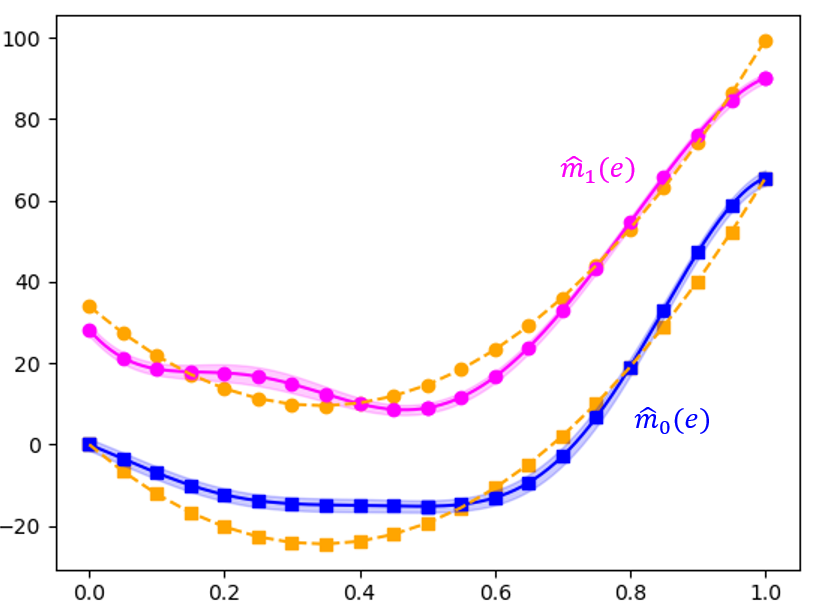}
}
\hfill
\subfloat[\FANCIFull: 7 Networks\label{fig:full-7}]{
    \includegraphics[width=0.30\linewidth]
    {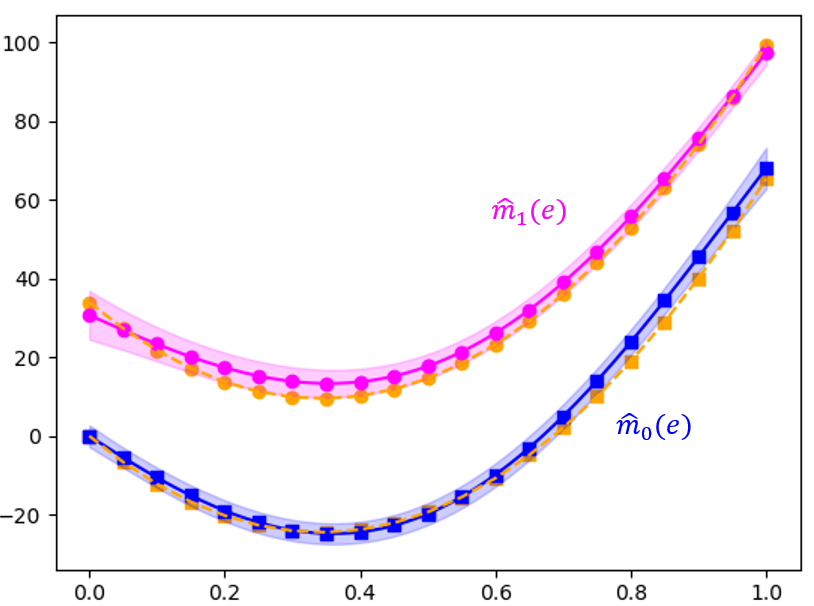}
}

% FANCI-Geo
\subfloat[\FANCIGeo: 3 Networks\label{fig:geo-3}]{
    \includegraphics[width=0.30\linewidth]
    {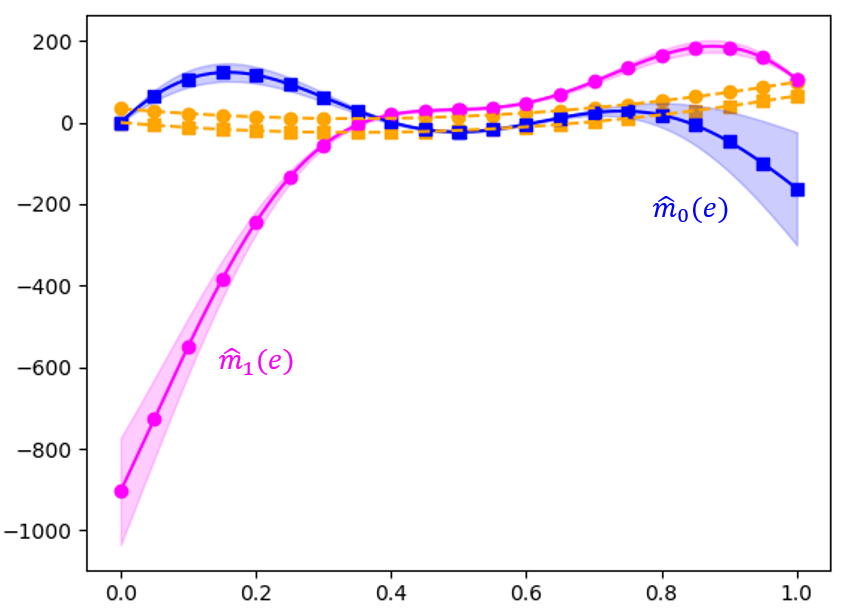}
}
\hfill
\subfloat[\FANCIGeo: 5 Networks\label{fig:geo-5}]{
    \includegraphics[width=0.30\linewidth]
    {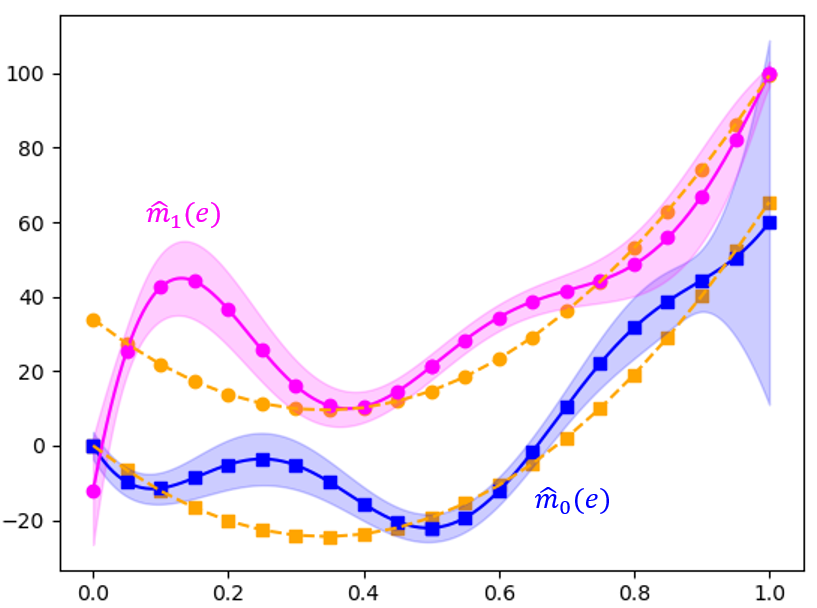}
}
\hfill
\subfloat[\FANCIGeo: 7 Networks\label{fig:geo-7}]{
    \includegraphics[width=0.30\linewidth]
    {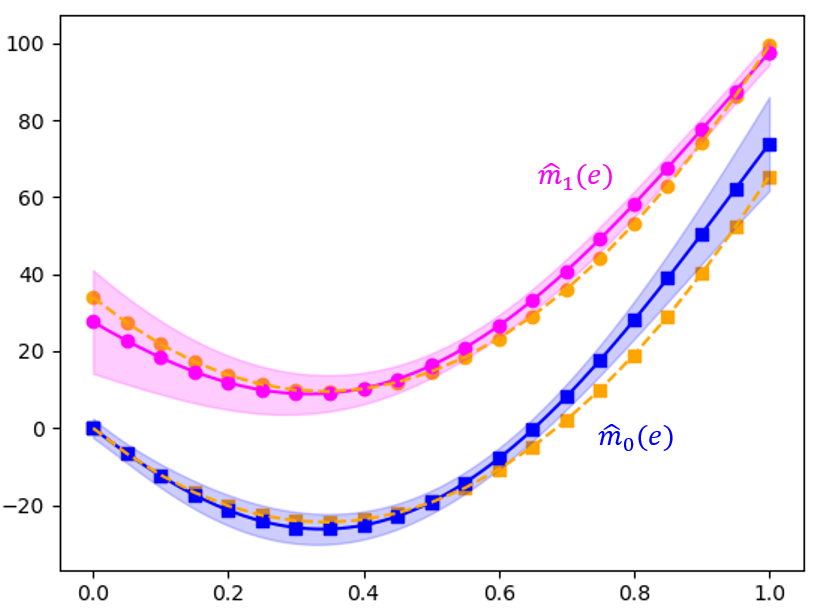}
}

% FANCI-Unc
\subfloat[\FANCIUnc: 3 Networks\label{fig:unc-3}]{
    \includegraphics[width=0.30\linewidth]
    {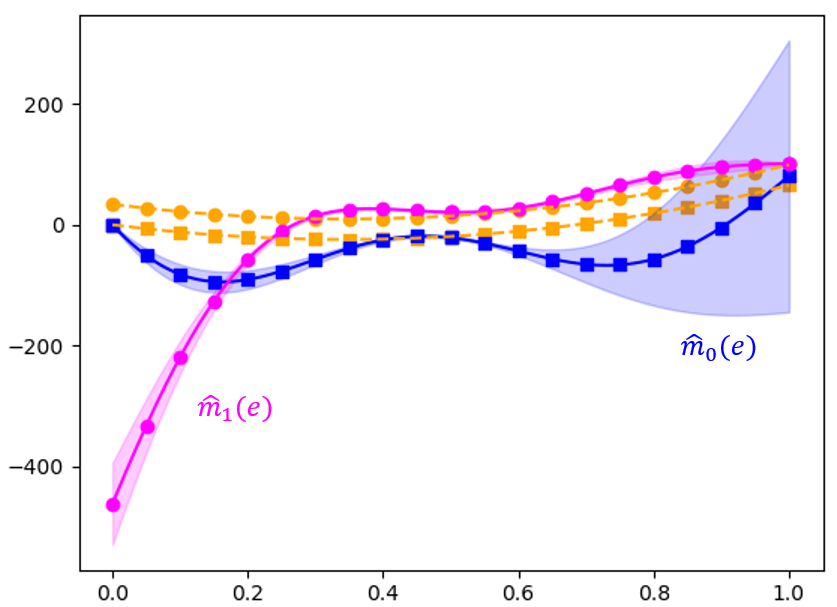}
}
\hfill
\subfloat[\FANCIUnc: 5 Networks\label{fig:unc-5}]{
    \includegraphics[width=0.30\linewidth]
    {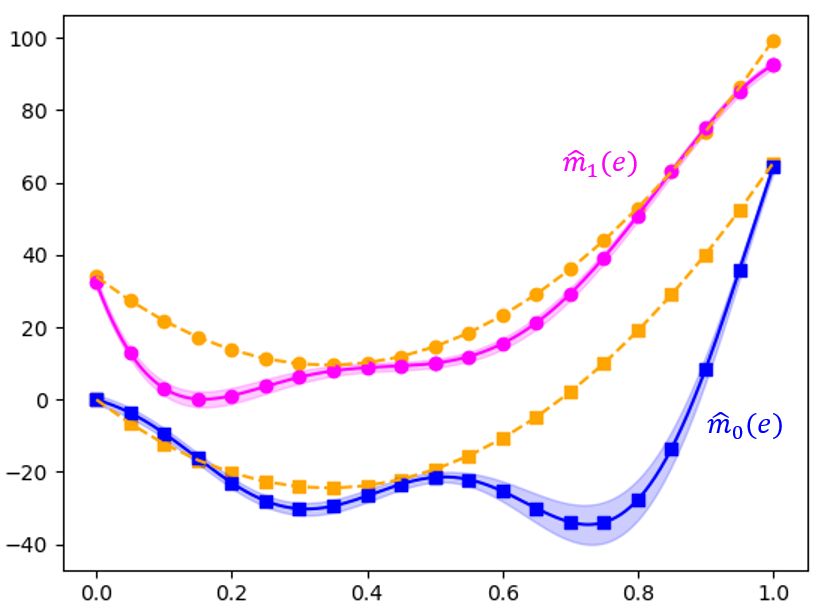}
}
\hfill
\subfloat[\FANCIUnc: 7 Networks\label{fig:unc-7}]{
    \includegraphics[width=0.30\linewidth]
    {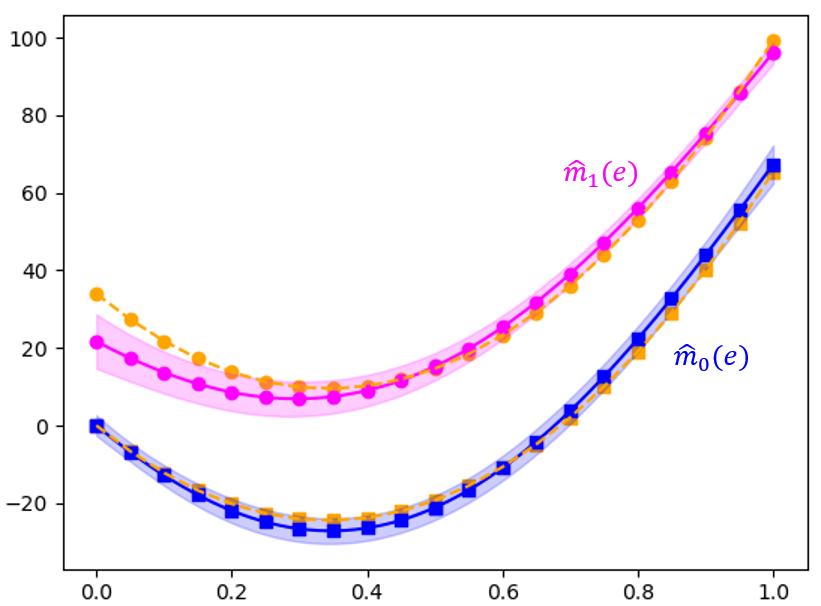}
}

% MMC
\subfloat[\MMC: 3 Networks\label{fig:mmc-3}]{
    \includegraphics[width=0.30\linewidth]
    {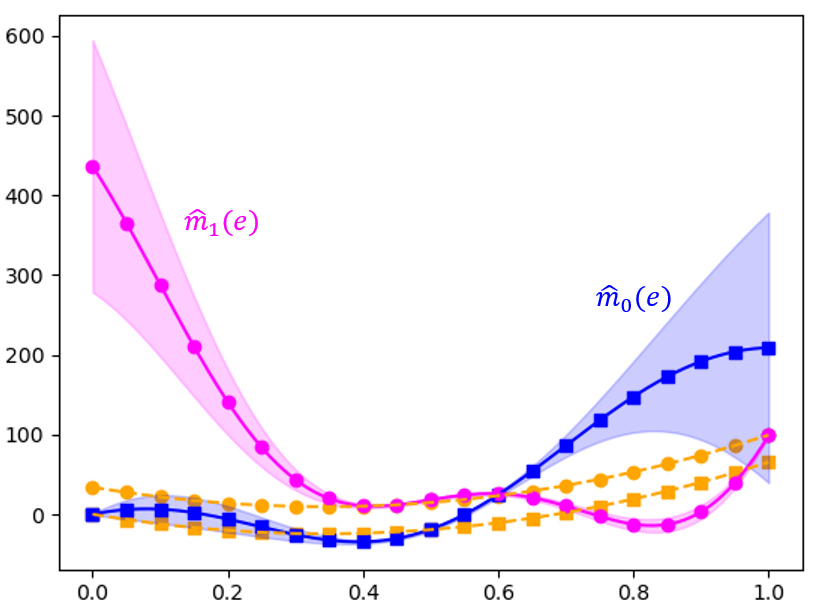}
}
\hfill
\subfloat[\MMC: 5 Networks\label{fig:mmc-5}]{
    \includegraphics[width=0.30\linewidth]
    {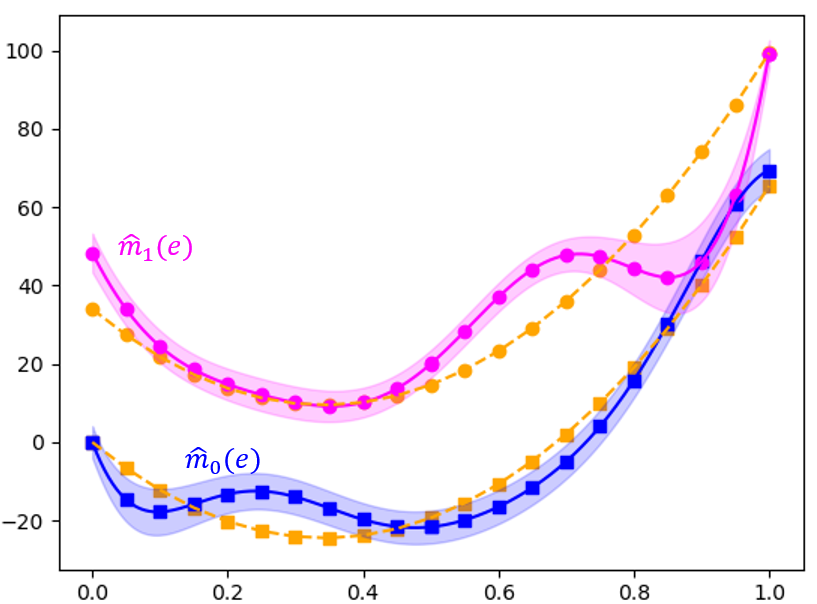}
}
\hfill
\subfloat[\MMC: 7 Networks\label{fig:mmc-7}]{
    \includegraphics[width=0.30\linewidth]
    {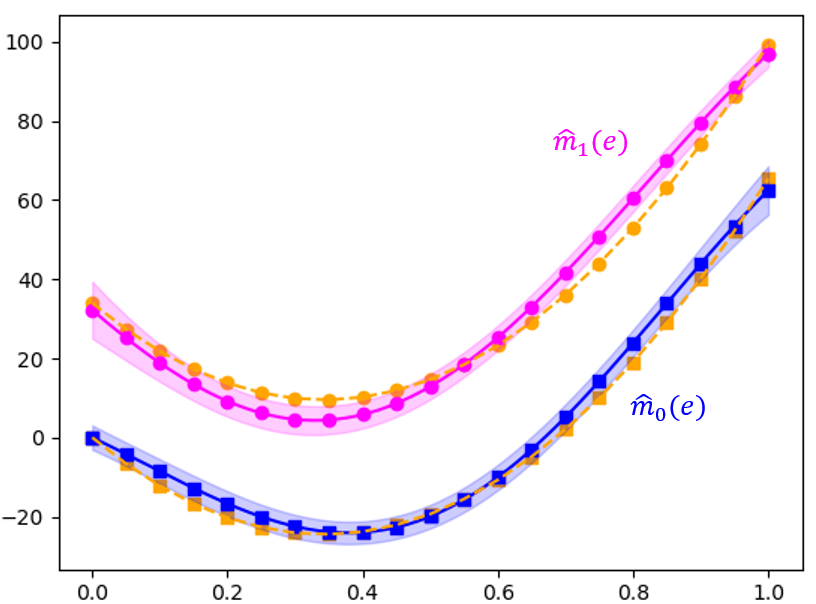}
}

% RAND
\subfloat[\RAND: 3 Networks\label{fig:rand-3}]{
    \includegraphics[width=0.30\linewidth]
    {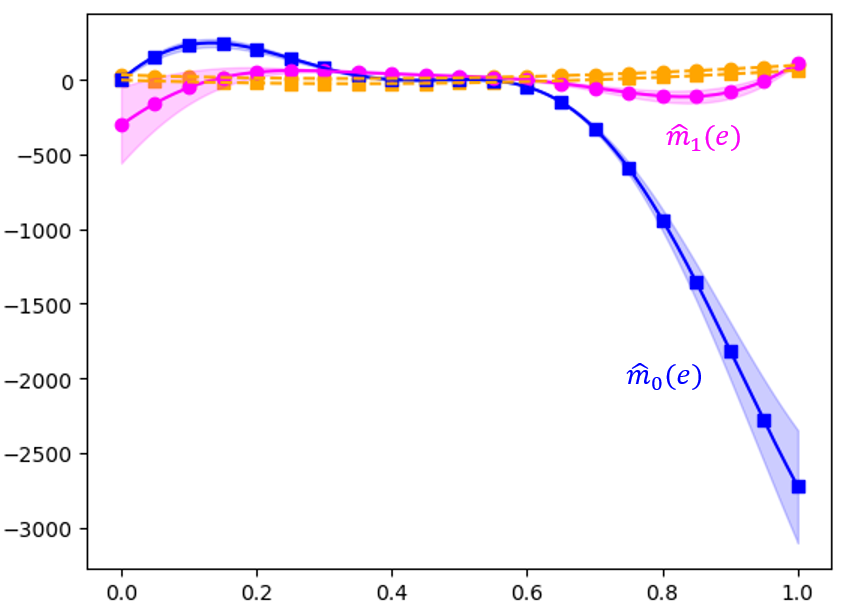}
}
\hfill
\subfloat[\RAND: 5 Networks\label{fig:rand-5}]{
    \includegraphics[width=0.30\linewidth]
    {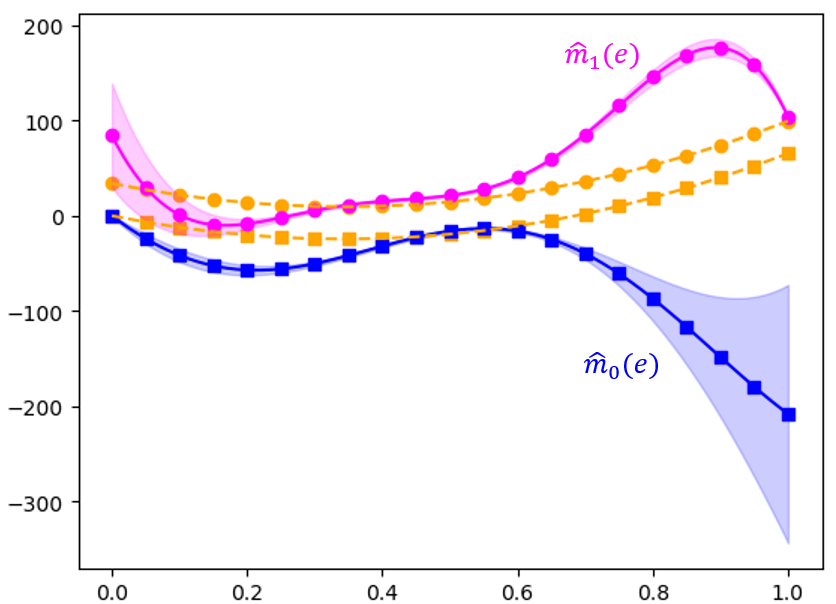}
}
\hfill
\subfloat[\RAND: 7 Networks\label{fig:rand-7}]{
    \includegraphics[width=0.30\linewidth]
    {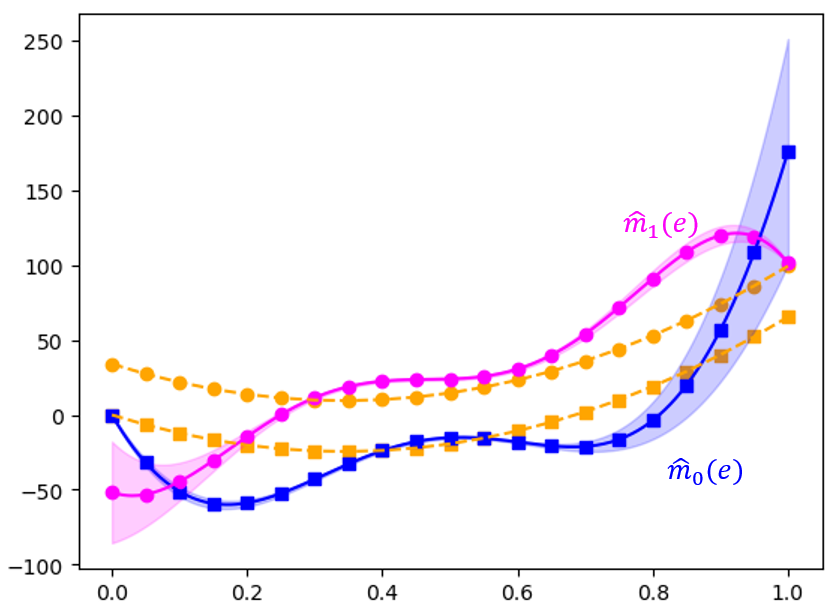}
}

\caption{Sequential Recovery of the Population-Average Response
Functions}
\label{fig:simulation-effect-comparison}
\end{figure}

Figure~\ref{fig:simulation-effect-comparison} presents the sequential
estimates of the population-average response functions \(\hat m_0(e)\) and
\(\hat m_1(e)\) as the number of network experiments increases in a
representative replication. The dashed curves denote the true response functions, the solid curves denote the corresponding posterior-mean estimates, and the shaded bands indicate the associated posterior uncertainty. Each row corresponds to one method, and
the columns report the estimates after \(3\), \(5\), and \(7\) network
experiments. The panels therefore illustrate both early-stage estimation
accuracy and the rate at which each acquisition rule improves
response-function recovery.

As shown in Figures~\ref{fig:full-3},
\ref{fig:full-5}, and \ref{fig:full-7}, \FANCIFull\ achieves the most
stable and balanced recovery of both population-average response
functions. \FANCIGeo\ and \FANCIUnc\ are somewhat less accurate after five
networks because target-based screening restricts the candidate search
space. By seven networks, however, this performance gap has narrowed
considerably, suggesting a limited empirical cost of screening in this setting. In particular, \FANCIGeo\
produces posterior means that more closely track the true response
functions, whereas \FANCIUnc\ yields narrower posterior uncertainty
bands, as shown in Figures~\ref{fig:geo-7} and \ref{fig:unc-7}. This
contrast accords with their respective emphasis on geometric coverage
and posterior uncertainty.

Among the benchmarks, \MMC\ recovers the main shapes of both response
functions and yields relatively narrow posterior uncertainty bands,
but its estimates are less stable than those of \FANCIFull\ in the
early stages of sequential acquisition, as seen by comparing
Figures~\ref{fig:mmc-5} and \ref{fig:full-5}. This difference arises
because \MMC\ prioritizes large GP model updates, which need not occur
in the regions where additional observations would most improve
global response-function recovery. \RAND\ performs worst among the
competing methods, with substantial bias and uncertainty remaining
near the exposure boundaries even after seven networks, as shown in
Figure~\ref{fig:rand-7}. This weaker performance stems from its
reliance on undirected sample accumulation: random assignments may
repeatedly generate common exposure patterns while providing limited
coverage of hard-to-realize regions.

\begin{table}[htbp]
\centering
\caption{Comparison of Causal Effect Function Learning Performance
Measured by MISE}
\label{tab:simulation-ise}
\renewcommand{\arraystretch}{1.10}
\small
\begin{tabular}{ccrrrrr}
\toprule
\multirow{2}{*}{Networks}
& \multirow{2}{*}{Effect}
& \multicolumn{3}{c}{Proposed Variants}
& \multicolumn{2}{c}{Benchmarks} \\
\cmidrule(lr){3-5}\cmidrule(lr){6-7}
& & \FANCIFull & \FANCIGeo & \FANCIUnc & \MMC & \RAND \\
\midrule
\multirow[c]{2}{*}{3}
& \(\tau_{\mathrm{overall}}\)
& \textbf{412.68} & 536.27 & 574.91 & 691.45 & 2486.73 \\
& \(\tau_{\mathrm{spill}}\)
& \textbf{348.52} & 429.36 & 401.78 & 527.94 & 3275.61 \\
\addlinespace[2pt]

\multirow[c]{2}{*}{5}
& \(\tau_{\mathrm{overall}}\)
& \textbf{48.73} & 71.84 & 79.26 & 108.57 & 914.38 \\
& \(\tau_{\mathrm{spill}}\)
& \textbf{36.95} & 55.42 & 49.18 & 72.63 & 1187.46 \\
\addlinespace[2pt]

\multirow[c]{2}{*}{7}
& \(\tau_{\mathrm{overall}}\)
& \textbf{8.64} & 11.37 & 13.52 & 19.86 & 382.71 \\
& \(\tau_{\mathrm{spill}}\)
& \textbf{7.26} & 10.14 & 8.83 & 14.57 & 274.39 \\
\bottomrule
\end{tabular}
\end{table}

Table~\ref{tab:simulation-ise} further provides a quantitative
comparison based on the MISE values for
\(\tau_{\mathrm{overall}}\) and \(\tau_{\mathrm{spill}}\). As
expected, \FANCIFull\ attains the smallest errors for both causal
effect functions after each reported number of network experiments,
with \FANCIGeo\ and \FANCIUnc\ generally ranking next. These results support the empirical benefit of combining GP learning with fill-distance coverage, while the theoretical bounds control approximation error rather than rank acquisition methods. In practice,
\FANCIFull\ is preferable when computational resources permit and
balanced recovery across the two estimands is the primary objective.
\FANCIGeo\ and \FANCIUnc\ provide less computationally demanding
alternatives, with \FANCIUnc\ particularly attractive when the
remaining posterior uncertainty is concentrated within a limited
portion of the target domain.

\subsection{Ablation Study}
\label{sec:simulation-ablation}

We conduct a model ablation to assess the importance of explicitly
modeling network interference by comparing interference-aware and
no-interference specifications. Specifically, the interference-aware
specification uses the augmented input
\(\boldsymbol U_{ki}=(E_{ki},\boldsymbol Z_{ki}^{\top})^{\top}\)
to estimate the conditional response surfaces
\(f_a(e,\boldsymbol z)\), which are standardized over the target
covariate distribution to obtain
\(m_a(e)=\int f_a(e,\boldsymbol z)\,
dF_{\boldsymbol Z}(\boldsymbol z)\), as described in
Section~\ref{sec:causal-framework}. In contrast, the no-interference
specification uses only the individual-level pretreatment covariates
\(\boldsymbol X_{ki}\), estimates treatment-specific conditional response
surfaces \(f_a(\boldsymbol x)\), and obtains the population-average
response
\(m_a=\int f_a(\boldsymbol x)\,
dF_{\boldsymbol X}(\boldsymbol x)\).

\begin{figure}[h]
\centering

\subfloat[No Interference: 2 Networks\label{fig:no-int-2}]{
    \includegraphics[width=0.30\linewidth]{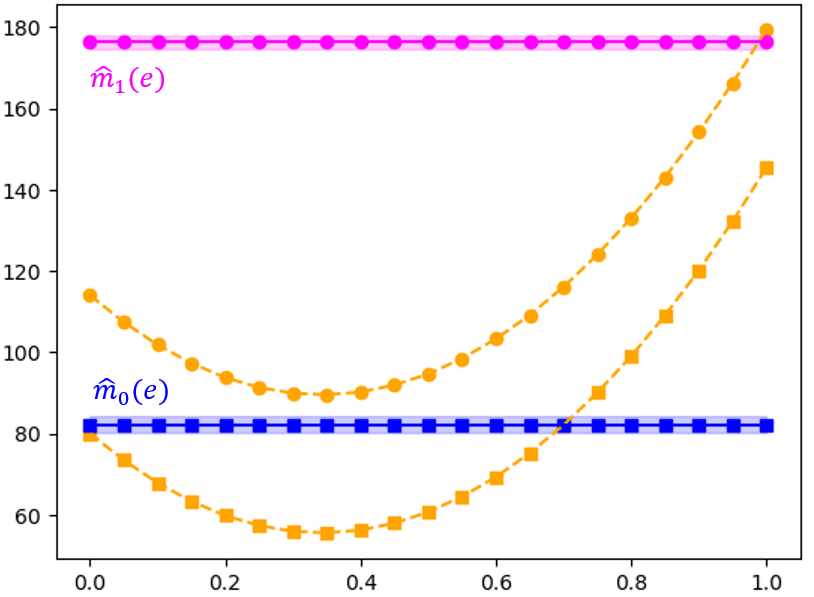}
}
\hfill
\subfloat[No Interference: 3 Networks\label{fig:no-int-3}]{
    \includegraphics[width=0.30\linewidth]{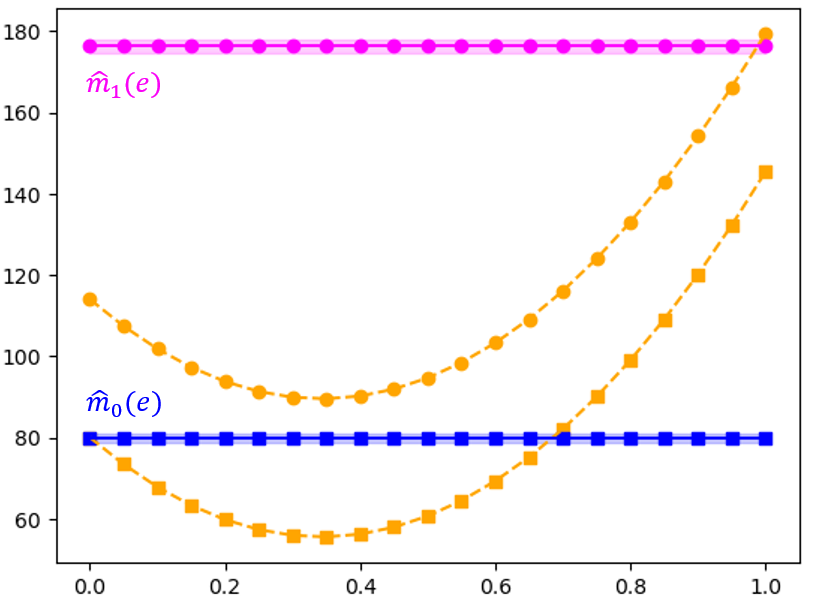}
}
\hfill
\subfloat[No Interference: 4 Networks\label{fig:no-int-4}]{
    \includegraphics[width=0.30\linewidth]{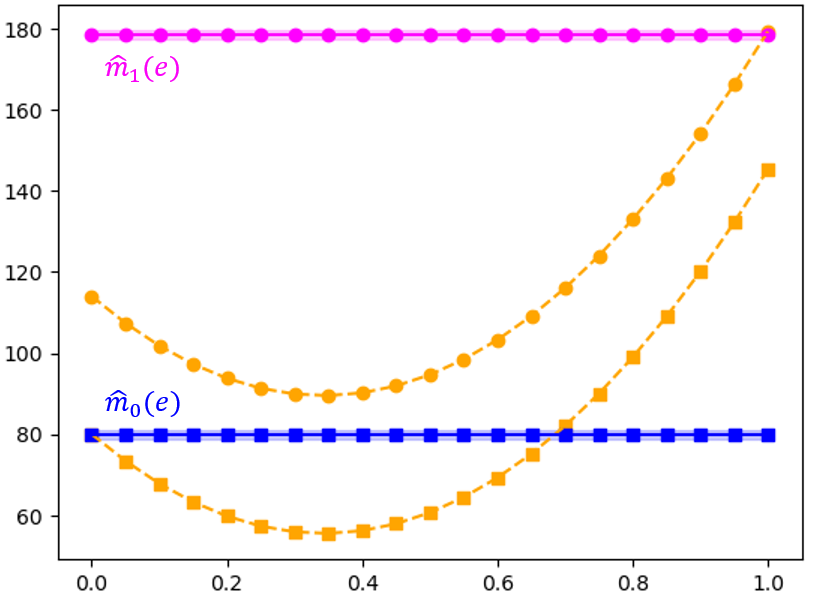}
}

\caption{Estimated Population-Average Response Functions Without Modeling
Neighborhood Exposure}
\label{fig:no-interference-ablation}
\end{figure}

Figure~\ref{fig:no-interference-ablation} presents the estimated
population-average responses under the no-interference specification
after two, three, and four network experiments. Unlike the
exposure-dependent estimates from \FANCIFull\ in
Figure~\ref{fig:full-7}, each no-interference estimate
\(\widehat m_a\) does not vary with \(e\); it reduces to a scalar rather
than a function and therefore appears as a horizontal line. Additional
data may improve estimation of the average treated and control responses,
but cannot recover the omitted dependence on neighborhood exposure.
This omission also carries over to the causal effect functions:
\(\widehat\tau_{\mathrm{spill}}(e)=0\) and
\(\widehat\tau_{\mathrm{dir}}(e)
=\widehat\tau_{\mathrm{overall}}(e)
=\widehat m_1-\widehat m_0\).
Thus, ignoring network interference removes the exposure-dependent
structure of the causal effects and leads to structural misspecification.

\subsection{Online-Game Application}
\label{sec:online-game}

We further evaluate \FANCI\ using data from an online experiment conducted by an online gaming company. For player \(i\) in network
\(k\), \(A_{ki}=1\) indicates receipt of the platform intervention,
whereas \(A_{ki}=0\) denotes the control condition. The outcome
\(Y_{ki}\) is the player's online time on the day of the experiment,
measured in minutes. The pretreatment covariates
\(\boldsymbol X_{ki}\) include the number of in-game friends and the
kill--death--assist (KDA) ratio, a standard measure of gaming
performance. Players are connected through in-game social networks,
and the neighborhood exposure \(E_{ki}\) is defined as the proportion
of player \(i\)'s friends who receive treatment, consistent with the
exposure mapping in
Section~\ref{sec:causal-framework-interference}. Our objective is to
learn how population-average engagement varies jointly with own
treatment and neighborhood exposure.

The target population consists of \(125{,}286\) players distributed
across 417 social networks. We sequentially experimented on 97 of
these networks, yielding observed outcomes for \(33{,}523\) players.
We implemented the coverage-guided target-screened strategy
\FANCIGeo\ described in Section~\ref{sec:solution-strategies}. All
five exposure--covariate coordinates were standardized before distance
evaluation, and the common diagonal weighting matrix
\(\boldsymbol M_0=\boldsymbol M_1\) assigns greater weight to the
exposure coordinate than to the four covariate coordinates. We set
\(\omega_0=\omega_1=0.5\) and retained the \(R=10\) networks with the
smallest coverage-based screening scores at each round for subsequent
assignment optimization.

\begin{figure}[htbp]
\centering

\subfloat[\FANCIGeo\ with Network Interference
\label{fig:case-geo}]{
    \includegraphics[width=0.4\linewidth]{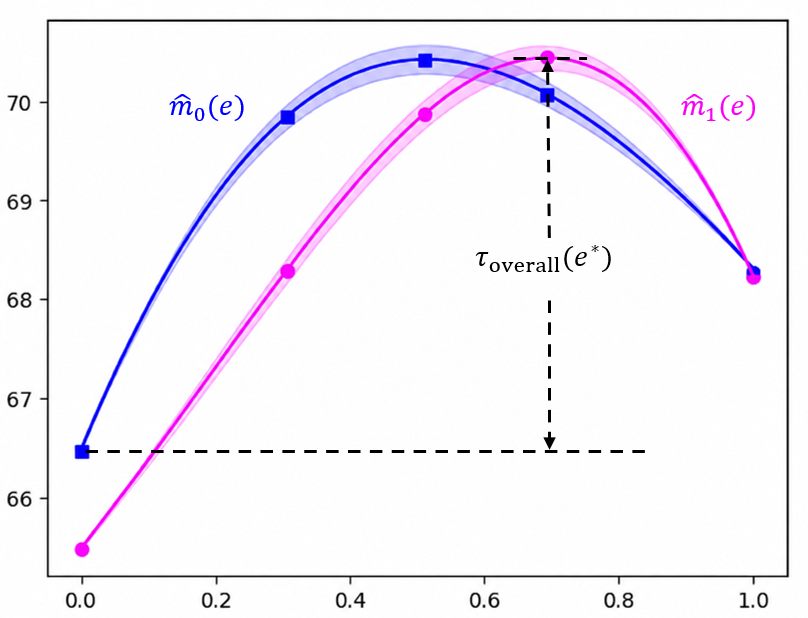}
}
\hspace{20pt}
\subfloat[Ignoring Network Interference
\label{fig:case-no-interference}]{
    \includegraphics[width=0.4\linewidth]{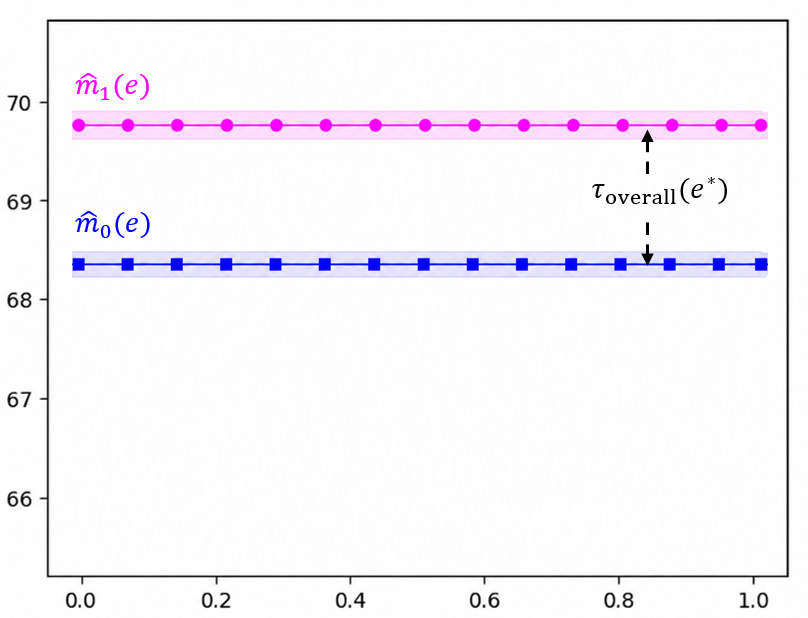}
}

\caption{Estimated Population-Average Response Functions in the
Online-Game Application}
\label{fig:case-study}
\end{figure}

Figure~\ref{fig:case-geo} presents the estimated population-average
response functions \(\widehat m_1(e)\) and \(\widehat m_0(e)\) obtained
by \FANCIGeo. Several findings emerge from the estimated response
functions:
\begin{enumerate}[label=(\roman*)]
    \item Both response functions exhibit a clear inverted-U pattern:
    player engagement initially increases with the proportion of treated
    friends but declines at high exposure levels. This pattern suggests
    that moderate neighborhood exposure is beneficial, whereas its marginal
    benefit diminishes as exposure becomes high.

    \item The direct effect,
    \(\widehat\tau_{\mathrm{dir}}(e)
    =\widehat m_1(e)-\widehat m_0(e)\), is relatively small over much of
    the exposure range. In particular, when few friends are treated,
    treating the player alone even provides negative benefit.
    Thus, the effectiveness of own treatment cannot be separated from
    the surrounding treatment environment.

    \item In contrast, the spillover effect,
    \(\widehat\tau_{\mathrm{spill}}(e)
    =\widehat m_0(e)-\widehat m_0(0)\), varies substantially with
    neighborhood exposure. The increase in engagement at moderate
    exposure levels indicates an important peer effect, consistent with
    reinforcement through treated friends.

    \item Combining own treatment and neighborhood exposure, the overall effect,
    \(\widehat\tau_{\mathrm{overall}}(e)
    =\widehat m_1(e)-\widehat m_0(0)\), reaches its maximum,
    \(\widehat\tau_{\mathrm{overall}}(e^\star)=3.971\), at
    \(e^\star=0.694\). Thus, among treated players, the largest estimated
    improvement relative to the untreated and unexposed baseline occurs
    when approximately \(69.4\%\) of their friends are treated. This
    finding suggests that the relevant decision is not simply how many
    players to treat, but how treatment should be allocated across the
    network to induce beneficial levels of neighborhood exposure. Note, however,
    that this pointwise optimum need not be feasible at the population
    level because neighborhood exposures are jointly determined by the
    network assignment.
\end{enumerate}

Figure~\ref{fig:case-no-interference} shows the result obtained when network
interference is ignored. Without neighborhood exposure in the model,
the two estimated population-average responses are constant in \(e\).
Because the estimated response under treatment exceeds that under
control, a decision based on the no-interference model alone would favor treating all players, which under
network interference corresponds to \((a,e)=(1,1)\). Evaluated using the interference-aware response estimate, this configuration has
\(\widehat\tau_{\mathrm{overall}}(1)=1.763\), below the
pointwise maximum of \(3.971\). Hence, ignoring network interference masks the nonlinear
dependence of treatment effectiveness on neighborhood exposure and can lead to
a materially different treatment decision. Taken together, these model-based results illustrate how \FANCI\ can recover
exposure-dependent treatment effects and inform
treatment decisions under network interference, without establishing a globally optimal feasible assignment.

%%%%%%%%%%%%%%%%%%%%%%%%%%%%%%%%%%%%%%%%%%%%%%%%%%%%%%%%%%%%%%%%%%%%%%
\section{Conclusion}
\label{sec:conclusion}
Network interference enriches the causal information of interest:
instead of a single scalar treatment effect, the causal effect may vary
with neighborhood exposure and become a function. This creates both new
opportunities for understanding heterogeneous treatment effects and new
challenges for learning them efficiently. Most existing studies focus
either on estimating prespecified scalar causal effects or on designing
experiments to improve their estimation, while sequential recovery of
the full population-average response function has received much less attention.
To fill this gap, this paper studies sequential recovery of the full
population-average response function under network interference with a limited experimental
budget. We
develop \FANCI, which uses treatment-specific Gaussian processes to
flexibly learn nonlinear conditional response surfaces and a fill-distance
criterion to guide new experiments toward poorly covered regions of the
target domain. Because feasible exposure patterns are induced jointly
by network structure and treatment assignment, we further develop
full-network and target-screened optimization strategies for selecting
the next experiment. The former directly optimizes the global coverage
objective, whereas the latter reduces computation by screening candidate
networks before optimization.

We establish theoretical guarantees that connect the proposed design
criterion to response-function recovery. In particular, we derive uniform
error bounds linking fill distance to estimation error for the response
and causal effect functions, and characterize the approximation loss
introduced by target screening. The numerical studies show that
\FANCI\ more accurately recovers nonlinear exposure-dependent effects
under limited experimental budgets, while the screened strategies offer
computationally cheaper alternatives to full-network optimization. In
the online-game application, accounting for neighborhood exposure reveals a
nonmonotone relationship between exposure and player
engagement, including an interior exposure level associated with the
largest estimated overall effect. Ignoring network interference obscures
this exposure-dependent structure and leads to substantially different
treatment implications.

Several extensions are worth pursuing. First, the assumption of a known
and stable network and exposure mapping can be relaxed to accommodate
network uncertainty, misspecification, or evolution during experimentation.
Second, larger networks and candidate pools motivate scalable GP
approximations and more structure-aware optimization methods. Third,
while \FANCI\ focuses on global response-function recovery, future work
may incorporate policy value, regret, or other decision-specific
objectives when the goal shifts from learning population-average response functions to
optimizing treatment deployment.

\bibliographystyle{informs2014}
\bibliography{references}

\begin{thebibliography}{4}
\providecommand{\natexlab}[1]{#1}
\providecommand{\url}[1]{\texttt{#1}}
\providecommand{\urlprefix}{URL }

\bibitem[{Gramacy(2020)}]{gramacy2020surrogates}
Gramacy RB (2020) \emph{Surrogates: {Gaussian} Process Modeling, Design, and
  Optimization for the Applied Sciences} (Boca Raton, FL: Chapman and
  Hall/CRC).

\bibitem[{Kanagawa et~al.(2018)Kanagawa, Hennig, Sejdinovic, \protect\BIBand{}
  Sriperumbudur}]{kanagawa2018gaussian}
Kanagawa M, Hennig P, Sejdinovic D, Sriperumbudur BK (2018) {Gaussian}
  processes and kernel methods: A review on connections and equivalences.
  \emph{arXiv preprint arXiv:1807.02582} .

\bibitem[{Rasmussen \protect\BIBand{} Williams(2006)}]{rasmussen2006gaussian}
Rasmussen CE, Williams CKI (2006) \emph{{Gaussian} Processes for Machine
  Learning} (Cambridge, MA: MIT Press).

\bibitem[{Wendland(2004)}]{wendland2004scattered}
Wendland H (2004) \emph{Scattered data approximation}, volume~17 (Cambridge
  University Press).

\end{thebibliography}


\begin{thebibliography}{44}
\providecommand{\natexlab}[1]{#1}
\providecommand{\url}[1]{\texttt{#1}}
\providecommand{\urlprefix}{URL }

\bibitem[{Alaeddini et~al.(2019)Alaeddini, Craft, Meka, \protect\BIBand{}
  Martinez}]{alaeddini2019sequential}
Alaeddini A, Craft E, Meka R, Martinez S (2019) Sequential {Laplacian}
  regularized {V}-optimal design of experiments for response surface modeling
  of expensive tests: An application in wind tunnel testing. \emph{IISE
  Transactions} 51(5):559--576.

\bibitem[{Aronow \protect\BIBand{} Samii(2017)}]{aronow2017estimating}
Aronow PM, Samii C (2017) Estimating average causal effects under general
  interference, with application to a social network experiment. \emph{The
  Annals of Applied Statistics} 11(4):1912--1947.

\bibitem[{Baird et~al.(2018)Baird, Bohren, McIntosh, \protect\BIBand{}
  {\"O}zler}]{baird2018optimal}
Baird S, Bohren JA, McIntosh C, {\"O}zler B (2018) Optimal design of
  experiments in the presence of interference. \emph{The Review of Economics
  and Statistics} 100(5):844--860.

\bibitem[{Bui et~al.(2026)Bui, Steiner, \protect\BIBand{}
  Stevens}]{bui2026genetic}
Bui T, Steiner SH, Stevens NT (2026) Genetic algorithm-based {Bayesian} optimal
  design for network experiments. \emph{Technometrics} 68(2):399--412.

\bibitem[{Chang et~al.(2021)Chang, Kim, Zhang, Pitt, \protect\BIBand{}
  Myung}]{chang2021data}
Chang J, Kim J, Zhang BT, Pitt MA, Myung JI (2021) Data-driven experimental
  design and model development using {Gaussian} process with active learning.
  \emph{Cognitive Psychology} 125:101360.

\bibitem[{Chokhachian et~al.(2026)Chokhachian, Katzfuss, \protect\BIBand{}
  Ding}]{chokhachian2026fastgp}
Chokhachian A, Katzfuss M, Ding Y (2026) Fast {Gaussian} process approximations
  for autocorrelated data. \emph{INFORMS Journal on Data Science}
  5(3):261--277.

\bibitem[{Eckles et~al.(2016)Eckles, Kizilcec, \protect\BIBand{}
  Bakshy}]{eckles2016estimating}
Eckles D, Kizilcec RF, Bakshy E (2016) Estimating peer effects in networks with
  peer encouragement designs. \emph{Proceedings of the National Academy of
  Sciences} 113(27):7316--7322.

\bibitem[{Erd{\H{o}}s \protect\BIBand{} R{\'e}nyi(1959)}]{erdos1959random}
Erd{\H{o}}s P, R{\'e}nyi A (1959) On random graphs. {I}. \emph{Publicationes
  Mathematicae Debrecen} 6:290--297.

\bibitem[{Fern{\'a}ndez-Lor{\'i}a \protect\BIBand{}
  Provost(2022)}]{fernandez2022causal}
Fern{\'a}ndez-Lor{\'i}a C, Provost F (2022) Causal decision making and causal
  effect estimation are not the same...and why it matters. \emph{INFORMS
  Journal on Data Science} 1(1):4--16.

\bibitem[{Forastiere et~al.(2021)Forastiere, Airoldi, \protect\BIBand{}
  Mealli}]{forastiere2021identification}
Forastiere L, Airoldi EM, Mealli F (2021) Identification and estimation of
  treatment and interference effects in observational studies on networks.
  \emph{Journal of the American Statistical Association} 116(534):901--918.

\bibitem[{Forastiere et~al.(2022)Forastiere, Mealli, Wu, \protect\BIBand{}
  Airoldi}]{forastiere2022estimating}
Forastiere L, Mealli F, Wu A, Airoldi EM (2022) Estimating causal effects under
  network interference with {Bayesian} generalized propensity scores.
  \emph{Journal of Machine Learning Research} 23(289):1--61.

\bibitem[{Goldberg(1989)}]{goldberg1989genetic}
Goldberg DE (1989) \emph{Genetic Algorithms in Search, Optimization, and
  Machine Learning} (Reading, MA: Addison-Wesley).

\bibitem[{Gramacy(2020)}]{gramacy2020surrogates}
Gramacy RB (2020) \emph{Surrogates: {Gaussian} Process Modeling, Design, and
  Optimization for the Applied Sciences} (Boca Raton, FL: Chapman and
  Hall/CRC).

\bibitem[{Hahn(1998)}]{hahn1998role}
Hahn J (1998) On the role of the propensity score in efficient semiparametric
  estimation of average treatment effects. \emph{Econometrica} 66(2):315--331.

\bibitem[{Halloran \protect\BIBand{} Struchiner(1995)}]{halloran1995causal}
Halloran ME, Struchiner CJ (1995) Causal inference in infectious diseases.
  \emph{Epidemiology} 6(2):142--151.

\bibitem[{Hirano et~al.(2003)Hirano, Imbens, \protect\BIBand{}
  Ridder}]{hirano2003efficient}
Hirano K, Imbens GW, Ridder G (2003) Efficient estimation of average treatment
  effects using the estimated propensity score. \emph{Econometrica}
  71(4):1161--1189.

\bibitem[{Hong \protect\BIBand{} Raudenbush(2006)}]{hong2006evaluating}
Hong G, Raudenbush SW (2006) Evaluating kindergarten retention policy: A case
  study of causal inference for multilevel observational data. \emph{Journal of
  the American Statistical Association} 101(475):901--910.

\bibitem[{Hu et~al.(2022)Hu, Li, \protect\BIBand{} Wager}]{hu2022average}
Hu Y, Li S, Wager S (2022) Average direct and indirect causal effects under
  interference. \emph{Biometrika} 109(4):1165--1172.

\bibitem[{Huang \protect\BIBand{} Pinson(2026)}]{huang2026active}
Huang X, Pinson P (2026) How to purchase labels? {A} cost-effective approach
  using active learning markets. \emph{INFORMS Journal on Data Science} 0(0).

\bibitem[{Hudgens \protect\BIBand{} Halloran(2008)}]{hudgens2008toward}
Hudgens MG, Halloran ME (2008) Toward causal inference with interference.
  \emph{Journal of the American Statistical Association} 103(482):832--842.

\bibitem[{Jagadeesan et~al.(2020)Jagadeesan, Pillai, \protect\BIBand{}
  Volfovsky}]{jagadeesan2020designs}
Jagadeesan R, Pillai NS, Volfovsky A (2020) Designs for estimating the
  treatment effect in networks with interference. \emph{The Annals of
  Statistics} 48(2):679--712.

\bibitem[{Kanagawa et~al.(2018)Kanagawa, Hennig, Sejdinovic, \protect\BIBand{}
  Sriperumbudur}]{kanagawa2018gaussian}
Kanagawa M, Hennig P, Sejdinovic D, Sriperumbudur BK (2018) {Gaussian}
  processes and kernel methods: A review on connections and equivalences.
  \emph{arXiv preprint arXiv:1807.02582} .

\bibitem[{Kosolwattana et~al.(2025)Kosolwattana, Wang, \protect\BIBand{}
  Lin}]{kosolwattana2025online}
Kosolwattana T, Wang H, Lin Y (2025) Online modeling and monitoring for
  dependent dynamic processes under resource constraints. \emph{INFORMS Journal
  on Data Science} 5(1):43--64.

\bibitem[{Li \protect\BIBand{} Wager(2022)}]{li2022random}
Li S, Wager S (2022) Random graph asymptotics for treatment effect estimation
  under network interference. \emph{The Annals of Statistics} 50(4):2334--2358.

\bibitem[{Li et~al.(2026)Li, Tian, Mo, \protect\BIBand{} Du}]{li2026gpbss}
Li Y, Tian Y, Mo H, Du S (2026) {Gaussian} process controlled {B}-spline
  surface. \emph{INFORMS Journal on Data Science} 0(0).

\bibitem[{Liu et~al.(2016)Liu, Hudgens, \protect\BIBand{}
  Becker-Dreps}]{liu2016inverse}
Liu L, Hudgens MG, Becker-Dreps S (2016) On inverse probability-weighted
  estimators in the presence of interference. \emph{Biometrika}
  103(4):829--842.

\bibitem[{Loeppky et~al.(2010)Loeppky, Moore, \protect\BIBand{}
  Williams}]{loeppky2010batch}
Loeppky JL, Moore LM, Williams BJ (2010) Batch sequential designs for computer
  experiments. \emph{Journal of Statistical Planning and Inference}
  140(6):1452--1464.

\bibitem[{Neyman(1990)}]{neyman1990application}
Neyman J (1990) On the application of probability theory to agricultural
  experiments: {Essay} on principles, {Section} 9. \emph{Statistical Science}
  5(4):465--472.

\bibitem[{Ogburn et~al.(2024)Ogburn, Sofrygin, D{\'i}az, \protect\BIBand{}
  van~der Laan}]{ogburn2022causal}
Ogburn EL, Sofrygin O, D{\'i}az I, van~der Laan MJ (2024) Causal inference for
  social network data. \emph{Journal of the American Statistical Association}
  119(545):597--611.

\bibitem[{Park \protect\BIBand{} Kang(2022)}]{park2022efficient}
Park C, Kang H (2022) Efficient semiparametric estimation of network treatment
  effects under partial interference. \emph{Biometrika} 109(4):1015--1031.

\bibitem[{Rasmussen \protect\BIBand{} Williams(2006)}]{rasmussen2006gaussian}
Rasmussen CE, Williams CKI (2006) \emph{{Gaussian} Processes for Machine
  Learning} (Cambridge, MA: MIT Press).

\bibitem[{Rosenbaum \protect\BIBand{} Rubin(1983)}]{rosenbaum1983central}
Rosenbaum PR, Rubin DB (1983) The central role of the propensity score in
  observational studies for causal effects. \emph{Biometrika} 70(1):41--55.

\bibitem[{Rubin(1974)}]{rubin1974estimating}
Rubin DB (1974) Estimating causal effects of treatments in randomized and
  nonrandomized studies. \emph{Journal of Educational Psychology}
  66(5):688--701.

\bibitem[{Rubin(1979)}]{rubin1979multivariate}
Rubin DB (1979) Using multivariate matched sampling and regression adjustment
  to control bias in observational studies. \emph{Journal of the American
  Statistical Association} 74(366a):318--328.

\bibitem[{Rubin(1980)}]{rubin1980randomization}
Rubin DB (1980) Randomization analysis of experimental data: The {Fisher}
  randomization test comment. \emph{Journal of the American Statistical
  Association} 75(371):591--593.

\bibitem[{Schulz et~al.(2018)Schulz, Speekenbrink, \protect\BIBand{}
  Krause}]{schulz2018tutorial}
Schulz E, Speekenbrink M, Krause A (2018) A tutorial on {Gaussian} process
  regression: Modelling, exploring, and exploiting functions. \emph{Journal of
  Mathematical Psychology} 85:1--16.

\bibitem[{Sobel(2006)}]{sobel2006randomized}
Sobel ME (2006) What do randomized studies of housing mobility demonstrate?
  causal inference in the face of interference. \emph{Journal of the American
  Statistical Association} 101(476):1398--1407.

\bibitem[{Ugander et~al.(2013)Ugander, Karrer, Backstrom, \protect\BIBand{}
  Kleinberg}]{ugander2013graph}
Ugander J, Karrer B, Backstrom L, Kleinberg J (2013) Graph cluster
  randomization: Network exposure to multiple universes. \emph{Proceedings of
  the 19th ACM SIGKDD International Conference on Knowledge Discovery and Data
  Mining}, 329--337 (ACM).

\bibitem[{Viviano(2026)}]{viviano2026experimental}
Viviano D (2026) Experimental design under network interference. \emph{Journal
  of Political Economy Microeconomics} Accepted for publication.

\bibitem[{Wendland(2004)}]{wendland2004scattered}
Wendland H (2004) \emph{Scattered data approximation}, volume~17 (Cambridge
  University Press).

\bibitem[{Zhang(2025)}]{zhang2025rerandomization}
Zhang Q (2025) Rerandomization algorithms for optimal designs of network {A/B}
  tests. \emph{Technometrics} 67(4):655--668.

\bibitem[{Zhang \protect\BIBand{} Kang(2022)}]{zhang2022locally}
Zhang Q, Kang L (2022) Locally optimal design for {A/B} tests in the presence
  of covariates and network dependence. \emph{Technometrics} 64(3):358--369.

\bibitem[{Zhang \protect\BIBand{} Wang(2025)}]{zhang2025online}
Zhang Z, Wang Z (2025) Online experimental design with estimation--regret
  trade-off under network interference. \emph{Advances in Neural Information
  Processing Systems}, volume~38.

\bibitem[{Zhao et~al.(2023)Zhao, Lin, Lin, \protect\BIBand{}
  Wu}]{zhao2023batch}
Zhao Y, Lin J, Lin J, Wu EQ (2023) Batch-mode active learning of {Gaussian}
  process regression with maximum model change. \emph{IEEE Transactions on
  Systems, Man, and Cybernetics: Systems} 53(12):7894--7900.

\end{thebibliography}

\end{document}

\title{Appendix for ``Fill-Distance-Guided Active Learning for Causal Function Recovery under Network Interference''}
\author{}
\date{}
\maketitle
\vspace{-2em}

\section{Gaussian-Process Specification and Estimation}
\label{app:gp}

For treatment level \(a\in\{0,1\}\), write the model input as
\(\boldsymbol u=(e,\boldsymbol z^\top)^\top\in\mathbb R^{h+1}\).
A concrete anisotropic specification consistent with the formulation in
the main paper is the squared-exponential kernel
\begin{equation}
\kappa_a(\boldsymbol u,\boldsymbol u')
=
\sigma_{\kappa,a}^2
\exp\!\left\{
-\frac{(e-e')^2}{2\ell_{a,e}^2}
-\frac{1}{2}
(\boldsymbol z-\boldsymbol z')^\top
\boldsymbol\Lambda_{a,z}^{-1}
(\boldsymbol z-\boldsymbol z')
\right\},
\label{eq:app-kernel}
\end{equation}
where \(\boldsymbol\Lambda_{a,z}\) is diagonal with positive entries.
The exposure coordinate and the covariate coordinates therefore have
separate length scales, allowing the response surface to vary at
different rates across input dimensions. Other stationary kernels, such
as Mat\'ern kernels, can be used without changing the estimation
procedure \citep{rasmussen2006gaussian,gramacy2020surrogates}.

Let \(\boldsymbol y_a^{(t)}\), \(\boldsymbol K_a^{(t)}\), and
\(\boldsymbol\mu_a^{(t)}\) be defined as in the main paper, and set
\(\boldsymbol C_a^{(t)}=\boldsymbol K_a^{(t)}+
\sigma_a^2\boldsymbol I\). For a parameter vector \(\boldsymbol\theta_a\)
containing the mean, kernel, and noise parameters, the treatment-specific
log marginal likelihood is
\begin{equation}
\begin{aligned}
\ell_a(\boldsymbol\theta_a)
={}&
-\frac{1}{2}
\bigl(\boldsymbol y_a^{(t)}-\boldsymbol\mu_a^{(t)}\bigr)^\top
\bigl(\boldsymbol C_a^{(t)}\bigr)^{-1}
\bigl(\boldsymbol y_a^{(t)}-\boldsymbol\mu_a^{(t)}\bigr)\\
&-\frac{1}{2}\log\det\boldsymbol C_a^{(t)}
-\frac{n_a^{(t)}}{2}\log(2\pi).
\end{aligned}
\label{eq:app-mll}
\end{equation}
The hyperparameters may be re-estimated after each experimental round by
maximizing \(\ell_a(\boldsymbol\theta_a)\), with positive scale
parameters optimized on the log scale. In the numerical implementation,
input standardization is fixed before sequential experimentation so that
changes in the fitted length scales are not confounded with changes in
coordinate scaling. The posterior mean and covariance then follow from
the standard Gaussian conditioning formulas reported in the main paper
\citep{rasmussen2006gaussian}.

The theoretical analysis treats the GP mean, kernel, and noise variance as fixed.
This separates the geometric approximation effect of the sequential
design from the estimation error due to noise and regularization.
Appendix~\ref{app:identification} verifies the fixed-parameter estimation bound under sequential sampling; additional error from re-estimating hyperparameters is outside this guarantee.

\section{Proofs}
\label{app:proofs}

\subsection{Identification and Discussion of Assumptions}
\label{app:identification}

We first establish the identification result used in the main paper.
Fix $a\in\{0,1\}$ and an attainable input $(e,\boldsymbol z)$ satisfying
positivity. Exchangeability (A2) gives
\begin{equation}
\mathbb E\!\left[Y_{ki}(a,e)\mid\boldsymbol Z_{ki}=\boldsymbol z\right]
=
\mathbb E\!\left[
Y_{ki}(a,e)\mid
A_{ki}=a,E_{ki}=e,\boldsymbol Z_{ki}=\boldsymbol z
\right].
\label{eq:app-identification-1}
\end{equation}
On the conditioning event on the right, consistency (A1) implies
$Y_{ki}=Y_{ki}(a,e)$, and hence
\begin{equation}
f_a(e,\boldsymbol z)
=
\mathbb E\!\left[
Y_{ki}\mid
A_{ki}=a,E_{ki}=e,\boldsymbol Z_{ki}=\boldsymbol z
\right].
\label{eq:app-identification-2}
\end{equation}
This identifies the response surface at supported configurations but
not, by itself, at unobserved exposure levels. For a fixed finite
network pool, inference at other inputs relies on model-based
interpolation or extrapolation. Along expanding network pools,
continuity and increasingly dense treatment-specific designs provide
the basis for recovery over a continuous target domain.

The preceding argument also clarifies the roles of Conditions
(A1)--(A4). The exposure mapping in (A1) must summarize the relevant
interference, while exchangeability (A2) rules out residual treatment
and exposure confounding after conditioning on pretreatment
covariates. In sequential experiments, the latter depends on the
network-selection and assignment mechanisms. Positivity (A3) requires
support under those mechanisms, not merely a feasible assignment, and
stability (A4) allows observations from different networks to identify
a common conditional response surface. These are causal assumptions
rather than consequences of the GP model.

Conditions (A5) and (A6) govern recovery beyond the observed inputs.
The target domain in (A5) must contain the inputs needed for
population standardization, although not every point is attainable
from a finite network pool. The RKHS membership in (A6) is a smoothness
assumption on the centered true response, rather than a consequence of
the working GP model. Its kernel-distance condition can, however, be
verified for the squared-exponential kernel in
Equation~\eqref{eq:app-kernel}. Let
$\boldsymbol D_a=\operatorname{diag}(\ell_{a,e}^{-2},
\boldsymbol\Lambda_{a,z}^{-1})$ and set
$R_a(s)=\exp(-s^2/2)$ and
$d_{\kappa,a}(\boldsymbol u,\boldsymbol v)
=\|\boldsymbol u-\boldsymbol v\|_{\boldsymbol D_a}$.
Since $\boldsymbol M_a$ is positive definite,
\begin{equation}
\begin{aligned}
d_{\kappa,a}(\boldsymbol u,\boldsymbol v)
&\le c_a\|\boldsymbol u-\boldsymbol v\|_{\boldsymbol M_a},\\
c_a
&=\left\{\lambda_{\max}\!\left(
\boldsymbol M_a^{-1/2}\boldsymbol D_a
\boldsymbol M_a^{-1/2}\right)\right\}^{1/2}<\infty.
\end{aligned}
\label{eq:app-kernel-compatibility}
\end{equation}
Thus the specified kernel satisfies the representation and metric
compatibility required in (A6).

We next verify the stochastic remainder in (A7) under adaptive
sampling with fixed GP parameters. At round $t$, for $a\in\{0,1\}$,
let $\lambda_a=\sigma_a^2>0$ and define
\begin{equation}
\begin{aligned}
\gamma_{a,t}
&=\frac12\log\det\!\left(
\boldsymbol I+\lambda_a^{-1}\boldsymbol K_a^{(t)}\right),\\
\nu_{a,t}
&=\sup_{\boldsymbol u\in\Omega_a}
\left\{\widehat\kappa_a^{(t)}
(\boldsymbol u,\boldsymbol u)\right\}^{1/2}.
\end{aligned}
\label{eq:app-uncertainty-quantities}
\end{equation}
Here $\gamma_{a,t}$ is the kernel information-gain quantity and
$\nu_{a,t}$ is the largest posterior standard deviation over the
target domain. For any $\delta\in(0,1)$, the following remainder yields the
joint probability statement in (A7):
\begin{equation}
\begin{aligned}
r_{a,t}(\delta)
=\nu_{a,t}\left[
B_a+\frac{\sigma_a}{\sqrt{\lambda_a}}
\sqrt{2\gamma_{a,t}+2\log(2/\delta)}\right].
\end{aligned}
\label{eq:app-remainder}
\end{equation}
As the inputs are selected adaptively, both $\gamma_{a,t}$ and
$\nu_{a,t}$, and hence the remainder, may be random.

To verify Equation~\eqref{eq:app-remainder}, let
$\mathcal T_a:\mathcal H_{\kappa_a}\to
\mathbb R^{n_a^{(t)}}$ be the evaluation operator at the observed
inputs, and write $g_a=f_a-\mu_a$ and
$\boldsymbol\varepsilon_a$ for the centered true response and the
observation-error vector. Let $\Pi_t$ be the orthogonal projection
onto the span of the observed kernel sections and set
$s_t=\mu_a+\Pi_tg_a$. With
$\mathcal V_a=\mathcal T_a^*\mathcal T_a+\lambda_a I$ and
$\mathcal T_ag_a=\mathcal T_a\Pi_tg_a$, the posterior mean satisfies
\begin{equation}
\widehat f_a^{(t)}-s_t
=-\lambda_a\mathcal V_a^{-1}\Pi_tg_a
+\mathcal V_a^{-1}\mathcal T_a^*\boldsymbol\varepsilon_a,
\label{eq:app-posterior-decomposition}
\end{equation}
The inputs of a network experiment are selected before its
outcomes are observed and can therefore be ordered predictably. Conditional
sub-Gaussianity in (A7) therefore permits the kernel
self-normalized martingale inequality
\begin{equation}
\left\|\mathcal T_a^*\boldsymbol\varepsilon_a
\right\|_{\mathcal V_a^{-1}}
\le\sigma_a
\sqrt{2\gamma_{a,t}+2\log(2/\delta)}
\label{eq:app-self-normalized}
\end{equation}
to hold for $a=0,1$ on a common event of probability at least
$1-\delta$, using a union bound across treatment levels. The
inequality follows from the Gaussian-mixture argument for
conditional sub-Gaussian exponential supermartingales and remains
valid for predictable, adaptively chosen inputs.
For $\kappa_{a,\boldsymbol u}
=\kappa_a(\boldsymbol u,\cdot)$, the posterior variance identity
$\widehat\kappa_a^{(t)}(\boldsymbol u,\boldsymbol u)
=\lambda_a\langle\kappa_{a,\boldsymbol u},
\mathcal V_a^{-1}\kappa_{a,\boldsymbol u}\rangle$
and $\mathcal V_a\succeq\lambda_a I$ then give, by
Cauchy--Schwarz and $\|g_a\|_{\mathcal H_{\kappa_a}}\le B_a$,
\begin{equation}
\begin{aligned}
&\left|\widehat f_a^{(t)}(\boldsymbol u)-
s_t(\boldsymbol u)\right|\\
&\qquad\le
\left[B_a+\frac{\sigma_a}{\sqrt{\lambda_a}}
\sqrt{2\gamma_{a,t}+2\log(2/\delta)}\right]
\left\{\widehat\kappa_a^{(t)}
(\boldsymbol u,\boldsymbol u)\right\}^{1/2}.
\end{aligned}
\label{eq:app-posterior-confidence}
\end{equation}
Taking the supremum over $\boldsymbol u\in\Omega_a$ in
Equation~\eqref{eq:app-posterior-confidence} yields
$\|\widehat f_a^{(t)}-s_t\|_{\infty,\Omega_a}
\le r_{a,t}(\delta)$, as required by (A7).

The bound also gives a sufficient condition for Corollary~1: for
each fixed $\delta$, $r_{a,t}(\delta)\to0$ in probability whenever
$(1+\gamma_{a,t})\nu_{a,t}^2\to0$ in probability, for both
treatment levels. Together with vanishing fill distances, this
implies the corollary's assumptions. Neither condition is guaranteed
by an arbitrary network pool or screening strategy.

Finally, (A8) holds when standardization is evaluated under the
known target distribution or over the complete finite target
population; otherwise, sampling or numerical-integration error must
be controlled separately.

\subsection{Proof of Theorem 1: Fill-Distance Error Bounds}
\label{app:proof-theorem1}

\begin{proof}
Fix \(a\in\{0,1\}\) and suppress the treatment index when no ambiguity
arises. Write \(g_a=f_a-\mu_a\in\mathcal H_{\kappa_a}\). Let
\(\mathcal H=\mathcal H_{\kappa_a}\), and let
\(\mathcal V_t\) be the linear span in \(\mathcal H\) of the kernel
sections \(\kappa_a(\boldsymbol v,\cdot)\) with
\(\boldsymbol v\in\mathcal U_a^{(t)}\). Denote by \(\Pi_t\) the
orthogonal projection onto \(\mathcal V_t\), and define the noise-free
kernel interpolant \(s_t=\mu_a+\Pi_tg_a\). The projection formulation remains
valid when the design contains repeated inputs and avoids any
invertibility requirement on a Gram matrix.

For \(\boldsymbol u\in\Omega_a\), define the power function
\begin{equation}
P_t(\boldsymbol u)
=
\left\|
\kappa_a(\boldsymbol u,\cdot)
-
\Pi_t\kappa_a(\boldsymbol u,\cdot)
\right\|_{\mathcal H}.
\label{eq:app-power}
\end{equation}
Because \(g_a-\Pi_tg_a\) is orthogonal to \(\mathcal V_t\), the reproducing
property and Cauchy--Schwarz inequality imply
\begin{equation}
\begin{aligned}
|f_a(\boldsymbol u)-s_t(\boldsymbol u)|
&=
\left|
\left\langle
g_a-\Pi_tg_a,
\kappa_a(\boldsymbol u,\cdot)
-
\Pi_t\kappa_a(\boldsymbol u,\cdot)
\right\rangle_{\mathcal H}
\right|\\
&\le
\|g_a-\Pi_tg_a\|_{\mathcal H}P_t(\boldsymbol u)
\le
B_a P_t(\boldsymbol u).
\end{aligned}
\label{eq:app-rkhs-error}
\end{equation}

Choose \(\boldsymbol v\in\mathcal U_a^{(t)}\) nearest to
\(\boldsymbol u\) in the \(\boldsymbol M_a\)-norm. Since projection
onto \(\mathcal V_t\) cannot be worse than approximation by the single
kernel section \(\kappa_a(\boldsymbol v,\cdot)\),
\begin{equation}
\begin{aligned}
P_t^2(\boldsymbol u)
&\le
\inf_{c\in\mathbb R}
\left\|
\kappa_a(\boldsymbol u,\cdot)
-c\kappa_a(\boldsymbol v,\cdot)
\right\|_{\mathcal H}^2\\
&=
\kappa_a(\boldsymbol u,\boldsymbol u)
-
\frac{\kappa_a(\boldsymbol u,\boldsymbol v)^2}
{\kappa_a(\boldsymbol v,\boldsymbol v)}.
\end{aligned}
\label{eq:app-power-nearest}
\end{equation}
Assumption~2 gives
\(\kappa_a(\boldsymbol u,\boldsymbol u)
=\kappa_a(\boldsymbol v,\boldsymbol v)=\sigma_{\kappa,a}^2\) and
\(\kappa_a(\boldsymbol u,\boldsymbol v)
=\sigma_{\kappa,a}^2
R_a(d_{\kappa,a}(\boldsymbol u,\boldsymbol v))\). Moreover,
\begin{equation}
 d_{\kappa,a}(\boldsymbol u,\boldsymbol v)
 \le
 c_a\|\boldsymbol u-\boldsymbol v\|_{\boldsymbol M_a}
 \le
 c_a\FD_a^{(t)}.
\label{eq:app-distance-control}
\end{equation}
Since \(R_a\) is nonincreasing and takes values in \([0,1]\),
Equations~\eqref{eq:app-rkhs-error}--\eqref{eq:app-distance-control}
yield
\begin{equation}
\|s_t-f_a\|_{\infty,\Omega_a}
\le
B_a\sigma_{\kappa,a}
\left\{
1-R_a^2(c_a\FD_a^{(t)})
\right\}^{1/2}
=
\rho_a(\FD_a^{(t)}).
\label{eq:app-deterministic-bound}
\end{equation}
This is the standard RKHS interpolation argument expressed directly in
terms of the fill distance \citep{wendland2004scattered,kanagawa2018gaussian}.

By Assumption~2(A7), with joint probability at least \(1-\delta\) for
\(a=0,1\),
\(\|\widehat f_a^{(t)}-s_t\|_{\infty,\Omega_a}
\le r_{a,t}(\delta)\). Combining this inequality with
Equation~\eqref{eq:app-deterministic-bound} gives
\begin{equation}
\left\|\widehat f_a^{(t)}-f_a\right\|_{\infty,\Omega_a}
\le
\rho_a(\FD_a^{(t)})+r_{a,t}(\delta)
=
\varepsilon_{a,t}(\delta).
\label{eq:app-response-bound}
\end{equation}
Continuity of \(R_a\) at zero and \(R_a(0)=1\) imply
\(\rho_a(h)\to0\) as \(h\downarrow0\).

It remains to propagate the response-surface bound through population
standardization. For every \(e\in\mathcal E\), Assumption~2(A5) ensures
that the inputs required by the target covariate distribution lie in
\(\Omega_a\). Therefore, on the same event,
\begin{equation}
\begin{aligned}
\left|\widehat m_a^{(t)}(e)-m_a(e)\right|
&=
\left|
\mathbb E_{\boldsymbol Z}
\left[
\widehat f_a^{(t)}(e,\boldsymbol Z)
-f_a(e,\boldsymbol Z)
\right]
\right|\\
&\le
\mathbb E_{\boldsymbol Z}
\left|
\widehat f_a^{(t)}(e,\boldsymbol Z)
-f_a(e,\boldsymbol Z)
\right|\\
&\le
\varepsilon_{a,t}(\delta).
\end{aligned}
\label{eq:app-standardization-bound}
\end{equation}
The same argument applies when the expectation is evaluated over the
complete finite target population. Taking the supremum over \(e\)
gives the asserted bound for \(\widehat m_a^{(t)}\).

Finally, the three causal-effect bounds follow from the triangle
inequality. For example,
\begin{equation}
\begin{aligned}
|\widehat\tau_{\mathrm{dir}}^{(t)}(e)-\tau_{\mathrm{dir}}(e)|
&\le
|\widehat m_1^{(t)}(e)-m_1(e)|
+
|\widehat m_0^{(t)}(e)-m_0(e)|,\\
|\widehat\tau_{\mathrm{spill}}^{(t)}(e)-\tau_{\mathrm{spill}}(e)|
&\le
|\widehat m_0^{(t)}(e)-m_0(e)|
+
|\widehat m_0^{(t)}(0)-m_0(0)|.
\end{aligned}
\label{eq:app-effect-bound}
\end{equation}
The overall-effect bound is identical to the direct-effect bound with
the second term evaluated at \(e=0\). Taking suprema completes the
proof.
\end{proof}

\subsection{Proof of Corollary 1: Uniform Consistency}
\label{app:proof-corollary1}

\begin{proof}
Fix \(a\in\{0,1\}\), \(\eta>0\), and \(\delta\in(0,1)\). By Theorem~1,
\begin{equation}
\begin{aligned}
&\Pr\!\left(
\|\widehat f_a^{(t)}-f_a\|_{\infty,\Omega_a}>\eta
\right)\\
&\qquad\le \delta+
\Pr\!\left(\rho_a(\FD_a^{(t)})+r_{a,t}(\delta)>\eta\right).
\end{aligned}
\end{equation}
The second probability tends to zero because both terms converge to zero in probability and \(\rho_a\) is continuous at zero. Thus the limit superior of the first probability is at most \(\delta\). Since \(\delta\) is arbitrary, \(\widehat f_a^{(t)}\) converges uniformly in probability to \(f_a\). The same argument applies to \(\widehat m_a^{(t)}\), and the causal-effect conclusions follow from the triangle-inequality bounds in Theorem~1.
\end{proof}

\subsection{Proof of Proposition 1: Target-Screening Approximation Bound}
\label{app:proof-proposition1}

\begin{proof}
Fix a network \(k\in\mathcal K_{t+1}\) and a screening assignment
\(\boldsymbol a_k\in\mathcal A_k^{\mathrm{scr}}\). For treatment level
\(a\), let
\begin{equation}
d_a(\boldsymbol a_k)
=
\min_{\substack{i\in\mathcal V_k:\ a_{ki}=a}}
\left\|
\boldsymbol u_{a,t+1}^{\dagger}-\boldsymbol u_{ki}
\right\|_{\boldsymbol M_a}.
\label{eq:app-target-mismatch}
\end{equation}
The screening construction guarantees that this minimum is well defined.
For any \(\boldsymbol v\in\Omega_a\), write
\begin{equation*}
D_a(\boldsymbol v)
=
\inf_{\boldsymbol u\in\mathcal U_a^{(t)}}
\|\boldsymbol v-\boldsymbol u\|_{\boldsymbol M_a}.
\end{equation*}
A point generated by \(\boldsymbol a_k\) lies within
\(d_a(\boldsymbol a_k)\) of the ideal target. Hence the triangle
inequality gives
\begin{equation}
\begin{aligned}
&\inf_{\boldsymbol u\in
\mathcal U_a^{(t+1)}(k,\boldsymbol a_k)}
\|\boldsymbol v-\boldsymbol u\|_{\boldsymbol M_a}\\
&\quad\le
\min\left\{
D_a(\boldsymbol v),
\|\boldsymbol v-\boldsymbol u_{a,t+1}^{\dagger}\|_{\boldsymbol M_a}
+d_a(\boldsymbol a_k)
\right\}\\
&\quad\le
\min\left\{
D_a(\boldsymbol v),
\|\boldsymbol v-\boldsymbol u_{a,t+1}^{\dagger}\|_{\boldsymbol M_a}
\right\}
+d_a(\boldsymbol a_k).
\end{aligned}
\label{eq:app-screening-triangle}
\end{equation}
Taking the supremum over \(\boldsymbol v\in\Omega_a\) shows that the
fill distance produced by \(\boldsymbol a_k\) is at most the fill
distance obtained by adding the ideal target plus
\(d_a(\boldsymbol a_k)\). Multiplying by \(\omega_a\), summing over
\(a\in\{0,1\}\), and then minimizing over
\(\boldsymbol a_k\in\mathcal A_k^{\mathrm{scr}}\) yields
\begin{equation}
\mathcal J_{t+1}^{\star}(k)
\le
\mathcal J_{t+1}^{\dagger}
+
\mathcal S_{t+1}(k),
\label{eq:app-network-bound}
\end{equation}
where we also used
\(\mathcal A_k^{\mathrm{scr}}\subseteq\mathcal A_k\).

Now let \(k^{\mathrm{scr}}\) be a network attaining
\(\min_{k\in\mathcal K_{t+1}}\mathcal S_{t+1}(k)\). Because the retained
set contains the \(R\ge1\) smallest screening scores,
\(k^{\mathrm{scr}}\in\widetilde{\mathcal K}_{t+1}\). Therefore,
\begin{equation}
\widetilde{\mathcal J}_{t+1}^{\star}
\le
\mathcal J_{t+1}^{\star}(k^{\mathrm{scr}})
\le
\mathcal J_{t+1}^{\dagger}
+
\min_{k\in\mathcal K_{t+1}}\mathcal S_{t+1}(k).
\label{eq:app-screened-upper}
\end{equation}
Since optimization over a subset cannot improve the full-network
optimum,
\(\mathcal J_{t+1}^{\star}
\le\widetilde{\mathcal J}_{t+1}^{\star}\). Subtracting
\(\mathcal J_{t+1}^{\star}\) from
Equation~\eqref{eq:app-screened-upper} proves the stated bound.
\end{proof}

\bibliographystyle{informs2014}
\bibliography{references}